\documentstyle[psfig,epsfig]{mn}

\newif\ifAMStwofonts

\ifoldfss
  \ifCUPmtlplainloaded \else
    \NewTextAlphabet{textbfit} {cmbxti10} {}
    \NewTextAlphabet{textbfss} {cmssbx10} {}
    \NewMathAlphabet{mathbfit} {cmbxti10} {} 
    \NewMathAlphabet{mathbfss} {cmssbx10} {} 
  \fi
  \ifAMStwofonts
    \ifCUPmtlplainloaded \else
      \NewSymbolFont{upmath} {eurm10}
      \NewSymbolFont{AMSa} {msam10}
      \NewMathSymbol{\upi}     {0}{upmath}{19}
      \NewMathSymbol{\umu}     {0}{upmath}{16}
      \NewMathSymbol{\upartial}{0}{upmath}{40}
      \NewMathSymbol{\leqslant}{3}{AMSa}{36}
      \NewMathSymbol{\geqslant}{3}{AMSa}{3E}

       \let\le=\leqslant
       \let\ge=\geqslant
    \fi
  \fi
\fi 

\ifnfssone
  \newmathalphabet{\mathit}
  \addtoversion{normal}{\mathit}{cmr}{m}{it}
  \addtoversion{bold}{\mathit}{cmr}{bx}{it}
  \newmathalphabet{\mathbfit} 
  \addtoversion{normal}{\mathbfit}{cmr}{bx}{it}
  \addtoversion{bold}{\mathbfit}{cmr}{bx}{it}
  \newmathalphabet{\mathbfss} 
  \addtoversion{normal}{\mathbfss}{cmss}{bx}{n}
  \addtoversion{bold}{\mathbfss}{cmss}{bx}{n}
  \ifAMStwofonts
    \ifCUPmtlplainloaded \else
      \UseAMStwoboldmath
      \makeatletter
      \new@mathgroup\upmath@group
      \define@mathgroup\mv@normal\upmath@group{eur}{m}{n}
      \define@mathgroup\mv@bold\upmath@group{eur}{b}{n}
      \edef\UPM{\hexnumber\upmath@group}
      \new@mathgroup\amsa@group
      \define@mathgroup\mv@normal\amsa@group{msa}{m}{n}
      \define@mathgroup\mv@bold\amsa@group{msa}{m}{n}
      \edef\AMSa{\hexnumber\amsa@group}
      \makeatother
      \mathchardef\upi="0\UPM19
      \mathchardef\umu="0\UPM16
      \mathchardef\upartial="0\UPM40
      \mathchardef\leqslant="3\AMSa36
      \mathchardef\geqslant="3\AMSa3E

       \let\le=\leqslant
       \let\ge=\geqslant
    \fi
  \fi
\fi 

\ifnfsstwo
  \DeclareMathAlphabet{\mathbfit}{OT1}{cmr}{bx}{it}
  \SetMathAlphabet\mathbfit{bold}{OT1}{cmr}{bx}{it}
  \DeclareMathAlphabet{\mathbfss}{OT1}{cmss}{bx}{n}
  \SetMathAlphabet\mathbfss{bold}{OT1}{cmss}{bx}{n}
  \ifAMStwofonts
    \ifCUPmtlplainloaded \else
      \DeclareSymbolFont{UPM}{U}{eur}{m}{n}
      \SetSymbolFont{UPM}{bold}{U}{eur}{b}{n}
      \DeclareSymbolFont{AMSa}{U}{msa}{m}{n}
      \DeclareMathSymbol{\upi}{0}{UPM}{"19}
      \DeclareMathSymbol{\umu}{0}{UPM}{"16}
      \DeclareMathSymbol{\upartial}{0}{UPM}{"40}
      \DeclareMathSymbol{\leqslant}{3}{AMSa}{"36}
      \DeclareMathSymbol{\geqslant}{3}{AMSa}{"3E}

       \let\le=\leqslant
       \let\ge=\geqslant
    \fi
  \fi
\fi 

\ifCUPmtlplainloaded \else
  \ifAMStwofonts \else 
    \def\upi{\pi}
    \def\umu{\mu}
    \def\upartial{\partial}
  \fi
\fi
\newcommand{\be}{\begin{equation}}
\newcommand{\ee}{\end{equation}}
\newcommand{\ba}{\begin{eqnarray}}
\newcommand{\ea}{\end{eqnarray}}
\newcommand{\nn}{\nonumber \\}
\newcommand{\Mpc}{\, h^{-1}{\rm Mpc}}
\newcommand{\r}{\mbox{\boldmath $r$}}
\newcommand{\rhat}{\hat{\r}}
\newcommand{\thetab}{\mbox{\boldmath $\theta$}}

\newcommand{\de}{\partial}
\newcommand{\rh}{\hat{r}}
\newcommand{\lgl}{\langle}
\newcommand{\rgl}{\rangle}
\newcommand{\nablab}{\mbox{\boldmath $\nabla$}}
\newcommand{\s}{\mbox{\boldmath $s$}}
\newcommand{\rhatb}{\mbox{\boldmath $\hat{r}$}}
\newcommand{\vb}{\mbox{\boldmath $v$}}
\newcommand{\Tr}{\mbox{\rm Tr}}
\newcommand{\lb}{\mbox{\boldmath $\ell$}}

\newcommand{\lbh}{\mbox{\boldmath $\hat{\ell}$}}

\title[Mapping the 3-D Dark Matter Potential] {Mapping the 3-D Dark
Matter Potential with Weak Shear } 
\author[D. Bacon \& A. N. Taylor] {D. J. Bacon\thanks{djb@roe.ac.uk} \& A. N. Taylor\thanks{ant@roe.ac.uk} \\ Institute for
Astronomy, Royal Observatory Edinburgh, Blackford Hill, Edinburgh, EH9
3HJ, U. K.}  \date{}

\pagerange{\pageref{firstpage}--\pageref{lastpage}}
\pubyear{2002}

\begin{document}

\maketitle

\label{firstpage}

\begin{abstract}

We investigate the practical implementation of Taylor's (2002)
3-dimensional gravitational potential reconstruction method using weak
gravitational lensing, together with the requisite reconstruction of
the lensing potential. This methodology calculates the 3-D
gravitational potential given a knowledge of shear estimates and
redshifts for a set of galaxies. We analytically estimate the noise
expected in the reconstructed gravitational field taking into account
the uncertainties associated with a finite survey, photometric
redshift uncertainty, redshift-space distortions, and multiple
scattering events. In order to implement this approach for future data
analysis, we simulate the lensing distortion fields due to various
mass distributions. We create catalogues of galaxies sampling this
distortion in three dimensions, with realistic spatial distribution
and intrinsic ellipticity for both ground-based and space-based
surveys. Using the resulting catalogues of galaxy position and shear,
we demonstrate that it is possible to reconstruct the lensing and
gravitational potentials with our method. For example, we demonstrate
that a typical ground-based shear survey with redshift limit $z=1$ and
photometric redshifts with error $\Delta z=0.05$ is directly able to
measure the 3-D gravitational potential for mass concentrations
$\ga10^{14} M_\odot$ between $0.1\la z\la0.5$, and can statistically
measure the potential at much lower mass limits. The intrinsic
ellipticity of objects is found to be a serious source of noise for
the gravitational potential, which can be overcome by Wiener filtering
or examining the potential statistically over many fields. We examine
the use of the 3-D lensing potential to measure mass and position of
clusters in 3-D, and to detect clusters behind clusters.

\end{abstract}

\begin{keywords}
Gravitation; Gravitational Lensing; Cosmology: Observations, Dark
Matter, Large-Scale Structure of Universe.
\end{keywords}

\section{Introduction}

Gravitational lensing affords us a direct method to probe the
distribution of matter in the universe, irrespective of its state or
nature. This deflection of light by the gravitational potential of
matter along its path can be observed as a local alteration of number
counts of background galaxies (magnification), or a distortion of
their shape (shear). It is the latter phenomenon which we will concern
ourselves with here; we will further restrict ourselves to the case
where this distortion is weak ($\la 10$\% change in the ellipticity of
the object). Despite the weakness of the effect, and the intrinsic,
nearly randomly orientated ellipticity of background galaxies, we can
measure the weak shear by averaging the ellipticity or shear estimates
of very many galaxies.

It has long been recognised that weak gravitational lensing is a
valuable tool for examining the two dimensional projected matter
distribution, and can consequently provide important information
regarding large-scale structure (see e.g. Bartelmann \& Schneider
2000, Bernardeau et al 1997, Jain \& Seljak 1997, Kaiser 1998). In
particular, the sensitivity of lensing to all the matter present,
including the dominant dark matter, ensures that the weak lensing
effect is an excellent probe for determining the quantity and
distribution of matter.

Weak lensing studies for a wide range of galaxy clusters have been
carried out, allowing precision measurements of the clusters' masses
and mass distributions (see e.g. Tyson et al 1990, Kaiser \& Squires
1993, Bonnet et al 1994, Squires et al 1996, Hoekstra et al 1998,
Luppino \& Kaiser 1997, Gray et al 2002). Moving to larger scales, the
shear due to large-scale structure has been accurately measured by
several groups (see e.g. van Waerbeke et al 2001, Hoekstra et al 2002, Bacon
et al 2002, Refregier et al 2002, Brown et al 2002). 

Redshift information has already been used in weak lensing studies, 
e.g. to determine the median redshifts of the lens and background
populations; most analyses then project the lensing information into a
2-dimensional projected mass distribution. Wittman et al (2001, 2002)
demonstrate the utility of using 3-D shear information by inferring
the redshift of a cluster using shear and photometric redshift
information for the galaxies in their sample. The importance of
including redshift information to remove intrinsic galaxy alignments
from shear studies has also been discussed (Heymans \& Heavens 2002,
King \& Schneider 2002a,b). Lens tomography has been studied as
a valuable means of introducing redshift information into shear power spectra
(e.g. Seljak 1998, Hu 1999, 2002, Huterer 2002, King \& Schneider
2002b), but only recently has the full reconstruction of the 3-D Dark
Matter distribution from lensing been considered (Taylor 2002, Hu \&
Keaton 2002).

In this paper, we seek to discuss a practical implementation for
reconstruction of the 3-D lensing and gravitational potentials from
weak lensing measurements and redshift information (whether
photometric or spectroscopic). We will use the reconstruction
procedure of Taylor (2002), which allows us to calculate the entire
3-D gravitational potential if we have a knowledge of shear estimates
and redshifts for a set of galaxies. This procedure is explained in
Section 2.

In Section 3 we discuss sources of uncertainty for our reconstruction,
examining analytically the effect of shot noise due to galaxy
ellipticity, the effects induced by a finite survey, photometric
redshift errors, redshift-space distortions, and multiple scatterings
of light rays.

We aim to test the reconstruction method using simulations of
realistic weak lensing data. In practice, for a particular volume of
space containing a mass distribution, the data which would result from
a real survey would be a set of galaxy ellipticities (in the weak
lensing regime, these will be dominated by intrinsic ellipticity with
a small gravitational lensing perturbation) and redshifts. The
ellipticities will typically be defined by the galaxies' quadrupole
moments (e.g. Kaiser et al 1995, Rhodes et al 2000) or estimated from
a decomposition of galaxy shape into eigenfunctions (Refregier \&
Bacon 2002, Bernstein \& Jarvis 2002). In order to provide a realistic
catalogue of these data, we simulate mass distributions, calculate the
expected lensing distortion in 3-D, and create a set of galaxies
probing this distortion with a realistic redshift distribution and
intrinsic ellipticity. We describe this procedure in detail in Section
4.1.

Given such a catalogue, we attempt to reconstruct the lensing
potential and gravitational potential (Section 4.2). We use a
generalised 3-D Kaiser-Squires (1993) inversion together with Taylor's
(2002) formalism to obtain the 3-dimensional gravitational potential
distribution.

In Section 5 we demonstrate the effectiveness of our implementation
in reconstructing lensing and gravitational potentials. We examine the
level of noise in our simulations in Section 6, including the Poisson
noise associated with having only a finite number of galaxies probing
the lensing distribution, and the noise from galaxies' intrinsic
ellipticities. We go on to examine the uncertainties caused by
photometric or spectroscopic redshift errors.

In Section 7 we study the utility of reconstructing only the 3-D
lensing potential without continuing to the gravitational potential;
this provides a useful method for detecting mass concentrations and
measuring their mass and 3-D position. Finally, we summarise our
results in Section 8.

\section{3-D Gravitational Potential Reconstruction}

In this section we will summarise the results of Taylor's (2002)
approach to reconstructing the 3-D gravitational potential, using weak
lensing measurements together with redshifts for all galaxies in the
lensing catalogue.

We begin by noting that the Newtonian gravitational potential $\Phi$
can be related to the density of matter $\rho$ by Poisson's equation,
\begin{equation}
\nabla^2 \Phi = 4 \pi G \rho_m \delta a^2 = \frac{3}{2}
\lambda_{H}^{-2} \Omega_m a^{-1} \delta, \label{pot-den}
\end{equation}
where we have introduced the cosmological scale factor $a$, the
density contrast $\delta = (\rho - \bar{\rho}) / \bar{\rho}$, the
Hubble length $\lambda_H=1/H_0 \approx 3000 \Mpc$, and the present-day
mass-density parameter $\Omega_m$.

We can study the impact of this gravitational potential on image
distortion due to gravitational lensing, by introducing the lensing
potential $\phi$. This is a measure of the distortion which is related
to the observable shear matrix by
\begin{equation}
\gamma_{ij} = \left(\de_i \de_j - \frac{1}{2} \delta^K_{ij}
\de^2\right)\phi,
\label{gij}
\end{equation}
where $\de_i \equiv r (\delta_{ij}- \rh_i \rh_j) \nabla_j = r(\nabla_i
- \rh_i \de_r)$ is a dimensionless, transverse differential operator,
and $\de^2 \equiv \de_i \de^i$ is the transverse Laplacian. The
indices $(i,j)$ take the values $(1,2)$, and we have assumed a flat
sky.

The lensing potential is also observable via the lens convergence
field
\be
    \kappa =\frac{1}{2} \de^2 \phi.
    \label{kappa}
\ee
The convergence field is related to the shear field by the
differential relation, first used by Kaiser \& Squires (1993);
\be
        \kappa = \de^{-2}\de_i \de_j \gamma_{ij},
        \label{ks}
\ee
where $\de^{-2}$ is the inverse 2-D Laplacian operator on a flat sky,
defined by
\be
        \de^{-2} \equiv \frac{1}{2 \pi}\int \! d^2 \!\theta \, \ln |\thetab - \thetab'|.
\ee
Note that transverse positions $\thetab$ in this equation will be
quoted in units of radians, so that the lensing quantities are
dimensionless. However, such angular positions can be arbitrarily
scaled, leading to simple scalings on the lensing quantities which we
will quote later.

A useful quantity for tracing noise and systematics in gravitational
lensing is the divergence-free field, $\beta$, defined by
\be
    \beta = \de^{-2} \varepsilon^n_{i} \de_{j} \de_n \gamma_{ij},
    \label{beta}
\ee
where $\epsilon^n_i = \left(\begin{array}{c c} 0 & -1 \\ 1 & 0
\end{array}\right)$.  If $\gamma_{ij}$ is generated purely by the
lensing potential, $\beta$ vanishes. But if there are
non-potential sources, due to noise, systematics or intrinsic
alignments, then $\beta$ will be non-zero. In addition, $\beta$ terms can
arise from finite fields, due to mode-mixing of shear fields
(e.g. Bunn 2002). We discuss this further in Section 3.2.

We can relate the lensing potential and the gravitational potential by
\begin{equation}
\phi(\r) = 2 \int_0^{r}\!\! dr' \, \left(\frac{r-r'}{r r'}\right)
\Phi(\r'),
\label{pot}
\end{equation}
in a spatially flat universe, with comoving distance $r$. This
equation assumes the Born approximation, in which the path of
integration is unperturbed by the lens.

Note that the lensing potential is really a 3-D quantity, although
usually it is regarded as a 2-D variable in the absence of redshift
information. In this case the usual practice is to average over the
redshift distribution of the source galaxies (e.g.  Bartelmann \&
Schneider 2001). It is in this 2-D approximation that the depth
information is lost in lensing.

Given that $\phi(\r)$ is really a 3-D variable, we can readily, and
exactly, invert equation (\ref{pot}) and recover the full 3-D
Newtonian potential (Taylor 2002);
\begin{equation}
\Phi(\r) = \frac{1}{2} \de_r r^2 \de_r \, \phi(\r)
\label{inv}
\end{equation}
where $\de_r = \rhat.\nabla$ is the radial derivative. Any lensing
field from real data will contain significant noise, and will thus
require smoothing if we are to perform this
differentiation. Interestingly it turns out that the lensing potential
obeys a second-order differential equation, which is $r^2$ times the
radial part of the 3-D Laplacian. This appears to be just a
coincidence given that the lensing kernel in equation (\ref{pot}) is
solely due to the geometric properties of the lens.

In order to reconstruct the gravitational potential, it is necessary
to find the lensing potential from the shear. This can be achieved by
using the Kaiser-Squires (1993) relation, generalised to 3-D:
\begin{equation}
    \widehat{\phi}(\r) = 2 \de^{-4} \de_i \de_j \ \gamma_{ij}(\r),
\label{kaiser-squires}
\end{equation}
where $\widehat{\phi}$ is an estimate of $\phi$. As the variance
of the shear field is formally infinite (Kaiser \& Squires 1993),
this distribution is usually binned and/or smoothed in the
transverse direction before calculating the lensing potential.

The above solution allows us to estimate the lensing potential only up
to an arbitrary function of the radial distance $r$:
\begin{equation}
\widehat{\phi}(\r) = \phi(\r) + \psi(r),
\end{equation}
where $\phi$ is the true lensing potential, and $\psi(r)$ is a
solution to
\begin{equation}
\left(\de_i \de_j - \frac{1}{2} \delta^K_{ij} \de^2\right)\psi =0 .
\end{equation}
This arbitrary radial behaviour is due to the fact that the shear only
defines the lensing potential up to a constant for each slice in
depth. However, we must tame this behaviour if we wish to apply
equation (\ref{inv}) to find the gravitational potential.

Fortunately, there are several opportunities for removing $\psi$.
Firstly, since $\langle \phi \rangle = 0$ for each slice in depth from
homogeneity and isotropy conditions, we can simply subtract $\langle
\widehat{\phi} \rangle$ from $\widehat{\phi}$ for each radial slice,
provided that our survey is large enough to discount errors in
$\langle \phi \rangle$. Thus
\begin{equation}
\Phi = \frac{1}{2} \de_r r^2  \de_r (\widehat{\phi} -\lgl
\widehat{\phi} \rgl). \label{unbiaspot}
\end{equation}
Alternatively, we can note that $\langle \de_r \phi \rangle = 0$ for
each radial slice, with a similar subtraction required.  Given these
procedures, it is clear that $\psi$ can be removed for large surveys,
where the necessary averages $\langle \tilde{\phi} \rangle$ or
$\langle \de_r \tilde{\phi} \rangle$ will have small uncertainty;
however for limited area surveys $\psi$ will not be estimated well,
leading to uncertainties on the reconstruction (see Section 3.1.2).

Similar relationships between the potential and the 3-D lens
convergence can also be written down:
\be
        \Phi = \de_r r^2 \de_r (\de^{-2} \kappa - \lgl \de^{-2} \kappa\rgl),
\ee
while the relationship between the matter density field and the 3-D
convergence is
\be
    \delta = \left(\frac{2 \lambda_H^2 a}{3\Omega_m}\right)
    \nabla^2 \de_r r^2 \de_r
        (\de^{-2} \kappa- \lgl \de^{-2} \kappa\rgl).
\ee
With these sets of equations, the 3-D lensing convergence, 3-D lensing
potential, 3-D Newtonian potential and 3-D matter density fields can
all be generated from combined shear and redshift information.

\section{The uncertainty in 3-D lensing}

\subsection{Shot-noise uncertainty in lensing fields}

Having written down the basic equations for the 3-D analysis of
gravitational lensing data, we now consider the various contributions
to the uncertainty in a reconstruction of these fields.

\subsubsection{The convergence field}

The covariance on a reconstructed, continuous convergence field due to
shot-noise is generally given by
\be
    \lgl \kappa(\r) \kappa(\r') \rgl_{SN} =
    \frac{\gamma^2_{\rm rms}}{n(\r)} \delta_D(\r-\r'),
\label{eqn:kerr}
\ee where $\gamma_{\rm rms}$ is the intrinsic dispersion of galaxy
shear estimates in one component (i.e. $\gamma_1$ or $\gamma_2$) due
to the non-circularity of galaxies, and $n(\r)$ is the observed space
density of galaxies in a survey.

In the case of a discretised map this reduces to
\be \lgl \kappa_i
\kappa_j \rgl_{SN} = \frac{\gamma^2_{\rm rms}}{N_{pix}} \delta^K_{ij},
\ee
where, for a constant 3-D number density of galaxies,
\be
N_{pix}= 3 \bar{n}\theta_{pix}^2 r^2 \Delta r / R^3
\ee 
is the 3-D pixel occupation number, $\bar{n}$ is the 3-D density of
galaxies, $\theta_{pix}$ is the pixel size, $\Delta r$ is the width of
the radial bins, $r$ is the radial distance, and $R$ is the chosen
limiting distance for the survey. We will compare this amplitude and
behaviour with our simulations in Section 6.

\subsubsection{The lensing potential field}

We now wish to describe the uncertainty expected for the lensing
potential field. We can write the covariance of the lensing
potential estimated from equation (\ref{kappa}) as
 \be
    \lgl \phi(\r) \phi(\r') \rgl_{SN} = 4
\partial^{-2}
 \partial'^{-2} \lgl \kappa(\r) \kappa(\r') \rgl_{SN}.  \label{phicov}
 \ee
We describe the procedure used to evaluate this covariance in the
Appendix, and here only quote the resulting variance of the lensing
potential at the centre of the survey ($\thetab=0$) as a function of
survey size, $\theta$, and radial position, $r$, in the flat-sky
limit:
 \be \lgl \Delta
 \phi^2(\thetab=0) \rgl_{SN} = \frac{5}{24 \pi}
 \frac{\gamma^2_{\rm rms}}{n(r)} \frac{\theta^2}{r^2} \delta_D(r-r').
 \label{transwind}
 \ee
Note that the uncertainty on the potential difference {\em increases}
with the survey area as a consequence of the 2-D flat-sky lensing
``force'' term, $\ln|\thetab-\thetab'|$, increasing with distance. We
will compare this behaviour with our simulations in Section 6.

Finally we can cast this in a more convenient form for
gravitational lensing in the discrete case:
 \be
    \lgl \Delta \phi^2(0) \rgl_{SN} =
    7.6 \times 10^{-16} \left( \frac{n_2}{30/[1']^2}\right)^{-1}
    \left( \frac{\theta}{1^\circ}\right)^2
    \left( \frac{R^3}{r^2 \Delta r}\right),
\label{phi_err}
\ee 
where we have assumed $\gamma_{\rm rms}=0.2$, $n_2 = \frac{1}{3} n
R^3$ is the 2-D total surface density of galaxies and $R$ is
the nominal depth of the survey.

In general the uncertainty in the 3-D lensing potential will also
depend on the pixel size as well as the total size of the survey.
This behaviour for the uncertainty on the lensing potential is now
due to the 2-D flat-sky lensing ``force'' term diverging at small
separation. Due to the complexity of analysing this effect
analytically we shall defer a more thorough treatment of studying
pixelisation effects with our simulations to Section 6.

\subsubsection{The Newtonian potential field}

We also wish to calculate the uncertainty on the 3-D Newtonian
potential. We can approach this, from equation (\ref{inv}), by
differentiating the lensing potential field leading in the
far-field limit to
 \be \lgl
  \Phi(\r) \Phi(\r') \rgl_{SN} = \frac{1}{4}r^2 r'^2 \partial_r^2
  \partial_r'^2 \lgl \phi(\r) \phi(\r') \rgl_{SN} .
 \ee
We again describe the detailed calculation of this quantity in the
Appendix, where we arrive at an expression for the variance on the
Newtonian potential:
 \be
    \lgl \Phi^2(r) \rgl_{SN} =  \frac{5}{64 \pi \sqrt{2 \pi}}
    \frac{\gamma^2_{\rm rms}}{r_{||}^3 n}\left( \frac{r}{r_{||}}\right)^4
    \left(\frac{r_{||}}{R} \right)^2 \theta^2 ,
    \label{eqn:phierr}
\ee
where $r_{||}$ is the smoothing radius for a radial Gaussian smoothing
of the field.  Hence we find that the shot-noise uncertainty is a
strong function of the radial smoothing, changing as the inverse fifth power
of the smoothing radius. We will again examine this behaviour in our
simulations in Section 6.

Finally, we can again cast this in a more convenient form for lensing:
\be
   \lgl \Phi^2(r) \rgl_{SN} =  1.1 \times 10^{-16}
    \left( \frac{n_2}{30/[1']^2}\right)^{-1}
    \left( \frac{\theta}{1^\circ}\right)^2
    \left( \frac{r}{r_{||}}\right)^4
    \left(\frac{R}{r_{||}} \right) .
\ee
Note that, as for the lensing potential, the error upon the
gravitational potential grows with survey or cell size.

The question of what survey area and radial smoothing to choose
leaves us with an optimisation problem. From equation
(\ref{eqn:phierr}) it would appear that we can reduce the
shot-noise by reducing the survey area, or increasing the radial
smoothing. However the latter will reduce the resolution of the
survey, and correspondingly lower the intrinsic signal, while the
former will increase the noise effects induced by a finite survey
area. We shall investigate further the effects arising from a
finite survey in Section 3.2, and comment further on the problem
of survey optimisation.

\subsubsection{The density field}

Finally the shot-noise uncertainty in the reconstructed density
field can be constructed from equation (\ref{pot-den});
 \be
    \lgl \delta(\r) \delta (\r') \rgl =
    \left(\frac{2 \lambda_H^2 a}{3 \Omega_m}\right)^2
    \nabla^2 \nabla'^2 \lgl
        \Phi(\r) \Phi(\r') \rgl.
    \label{delerr}
\ee In the far-field approximation the Laplacian can be written
$\nabla^2= (\de_r^2 + R^{-2} \de^2)$. In general equation
(\ref{delerr}) must be calculated numerically, but for points
along the centre of the survey the shot-noise contribution to the
variance of the density field can be calculated analytically,
reducing to
 \ba
    \lgl \delta^2(r) \rgl_{SN} &=&
   \left(\frac{2}{9 \pi^3}\right)^{\frac{1}{2}}
    \left(\frac{a}{\Omega_m}\right)^2
            \frac{\gamma_{\rm rms}^2}{n r_{||}^3}
        \left( \frac{\lambda_H}{R}\right)^4
    \left( \frac{r^2}{r_{||}\theta R}\right)^2 \nn
    & & \times
                \left[1+\frac{175}{24}
                \left(\frac{\theta R}{r_{||}} \right)^4 \right].
    \label{delerr2}
 \ea
 If we set $a=1/2$, and $\gamma_{\rm rms}=0.2$, and again
 define $n_2=\frac{1}{3}nR^3$ as the surface galaxy density, the variance
 of the density field  can be expressed in the form
 \ba
     \lgl \delta^2(r) \rgl_{SN} &=& 2.9 \times 10^{-8}
    \left( \frac{\Omega_m}{0.3}\right)^{-2}
    \left( \frac{n_2}{[30/1']^2}\right)^{-1}
    \left( \frac{\theta}{1^\circ}\right)^{-2} \nn
         \!\!\!&  \times & \!\!\!\!\!\!
            \left( \frac{\lambda_H}{r_{||}}\right)^4
    \left( \frac{r^4}{r_{||} R^3}\right)
    \left[1+6.8 \times 10^{-7}
                \left(\frac{(\theta/1^\circ) R}{r_{||}} \right)^4
                \right], \nn
 \ea

  To illustrate this we find that the variance in the
reconstructed density field for $r_{||}=0.1$, $R=1$ and $r=1$,
where we express distances here in redshift for a Euclidean
universe, with $n_2=30$ galaxies per sq. arcmin. and
$\Omega_m=0.3$ is given by 
\be 
\lgl \delta^2(r) \rgl_{SN} = 0.003 \left( \frac{\theta}{1^\circ}\right)^{-2}
 \left[1+ 0.007 \left( \frac{\theta}{1^\circ}\right)^{4} \right],
\label{delfinerr}
\ee
which has a minimum at $\theta=3.2^\circ$ of $\lgl \delta^2
\rgl^{1/2}=0.02$.  Since we expect the amplitude of density
perturbations on these scales, $\lambda \approx 100 \Mpc$, to be
smaller than this, $\sigma_{\delta} \approx 0.01$, we can expect that
filtering (e.g. Wiener filtering, see Section 3.3) will be
required to extract a large-scale map of the 3-D density field, even
at the scale which minimises noise.

\subsection{Uncertainty and mode-mixing due to finite fields}

\subsubsection{Finite surveys}

As well as shot-noise arising from the discrete sampling of the
shear field by the survey galaxies, there is an additional
uncertainty in the reconstruction of the density field for finite
area surveys due to the reconstruction process being nonlocal (via
the inverse Laplacian). The nonlocal behaviour of density
reconstruction over a finite survey area also gives rise to a
mixing of modes. Here we try to quantify for the first time for
lensing the effects of mode-mixing and reconstruction noise
arising from finite survey areas. The following analysis will be
applicable to either 3-D or 2-D lensing studies, as the results will
be true for either a series of redshift slices or an overall
2-dimensional projection.

We may calculate the effect of a finite shear sample by multiplying
the shear field by an arbitrary window function;
\be
        \gamma'_{ij}(\r) = W(\r) \gamma_{ij}(\r).
 \label{wind}
\ee The lens convergence field can be estimated via equation
(\ref{ks}), but since here there is only a finite field to
integrate over, the effect of the window function is to truncate
the effects due to the distant shear field and induce mode-mixing
between $\kappa$ and $\beta$.

If we expand the shear in Fourier modes on the flat sky, the observed
shear is a convolution of the intrinsic shear and the window
function. For convenience we shall use continuous transforms, although
for a finite sky a discrete Fourier transform with suitable boundary
conditions is more practical. A scalar quantity $f(\thetab,r)$ can be
expanded in a 2-D Fourier series by 
 \be
    f(\thetab,r) = \int \! \frac{d^2 \ell}{(2 \pi)^2} \,
        f(\lb,r) e^{i \lb . \thetab }.
 \ee
  Decomposing the observed shear matrix given by equation
(\ref{wind}) into the Fourier decomposed $\kappa(\lb)$ and
$\beta(\lb)$ fields using equations (\ref{ks}) and (\ref{beta}),
and Fourier transforming again we find the relationships between
the reconstructed and true $\kappa$- and $\beta$-modes are \ba
    \kappa'(\lb) \!\!\!\!&=&\!\!\!\! \int \!\frac{d^2 \ell'}{(2 \pi)^2} \,W(\lb-\lb')[
         \kappa(\lb') \cos 2 \varphi  -  \beta(\lb') \sin 2 \varphi], \nn
    \beta'(\lb) \!\!\!\!&=& \!\!\!\!\int \!\frac{d^2 \ell'}{(2 \pi)^2}\, W(\lb-\lb')[
         \beta(\lb') \cos 2 \varphi +  \kappa(\lb') \sin 2 \varphi],
\ea where $\cos \varphi = \lbh .\lbh'$. Hence we see that a finite
survey will give rise to a spurious $\beta$-field. Here we have
suppressed the radial dependence for clarity, so these equations
are directly applicable to reconstruction of the convergence field
in 2-D lensing.

If we assume negligible intrinsic $\beta$ fields then the real-space
convergence is given by
\be
    \kappa'(\thetab) = \int\! \frac{d^2 \ell}{(2 \pi)^2} \kappa (\lb)
    \widetilde{W}_\ell(\thetab) e^{-\ell^2 \theta_s^2/2},
\ee
where
 \be
    \widetilde{W}_\ell(\thetab) = \int \! \frac{d^2 \ell'}{(2 \pi)^2} W(\lb -
    \lb') \cos 2 \varphi  \, e^{i\lb'.\thetab}
 \ee
 is the total effect of the finite survey area, and we have
 assumed the shear field is Gaussian smoothed on a scale
 $\theta_s$.

\subsubsection{Variance of the convergence field}

The variance measured in the finite-survey convergence field on a smoothing
scale of $\theta_s$ is given by
 \be
    \lgl \kappa'^2(\thetab) \rgl = \int \! \frac{d^2 \ell}{(2 \pi)^2}
    C_{\ell}^{\kappa \kappa} |\widetilde{W}_\ell(\thetab) |^2
    e^{\ell^2 \theta_s^2},
\ee
where
\be
    \lgl X(\lb) Y(\lb) \rgl = (2\pi)^2 C^{XY}_\ell
    \delta_D(\lb-\lb')
\ee
for the isotropic fields $X(\thetab)$ and $Y(\thetab)$.

If we approximate the window function by a Gaussian of radius
$\theta$ with Fourier transform
 \be W(\ell) = 2 \pi \theta^2 \exp(- \ell^2/2 \theta^2),
 \ee
 we can evaluate $\widetilde{W}_\ell(\thetab)$
 at $\thetab=0$, yielding
 \be \widetilde{W}_\ell(0) = 1+ \frac{ 2(
 e^{-\ell^2 \theta^2/2}-1)}{\ell^2 \theta^2}.
 \ee
\begin{figure}
\centering
\begin{picture}(200,200)
\includegraphics{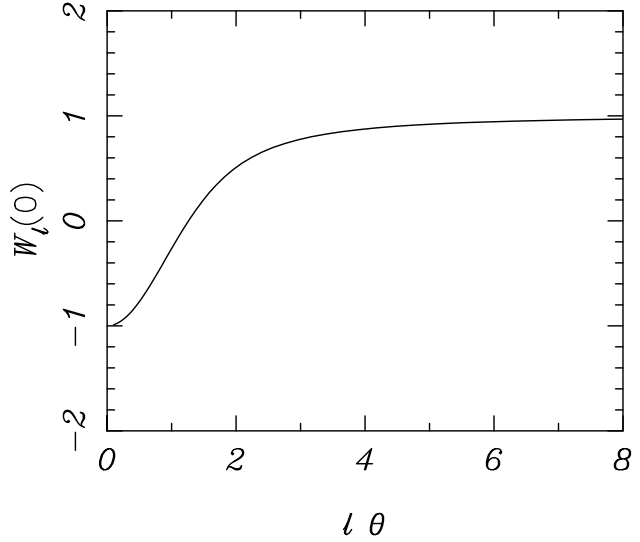}
\end{picture}
\caption{The finite-field lensing window function,
$\widetilde{W}_\ell(0)$, for a Gaussian window of radius $\theta$,
calculated at the centre of the survey as a function of $\ell
\theta$. } \label{window}
\end{figure}
We plot this window function, giving the contribution of convergence
modes to the observed convergence, in Figure \ref{window}. The general
effect of the window function is to act as a high-pass filter. At
$\ell \sim 1/\theta$ all the convergence modes are destroyed, while at
large $\ell \theta$ the window function tends to unity as all the
modes on scales below the scale of the survey contribute to the
variance. Interestingly at low $\ell \theta$ this function goes
negative, indicating that modes larger than the survey area are
heavily distorted.

For the Gaussian window function, no $\beta$-modes are generated
at the centre of the field due to the symmetry of the window.
Hence the $\beta$-modes that are generated by the finite window
are distributed nonlocally over the survey area.

\begin{figure}
\centering
\begin{picture}(200,200)
\includegraphics{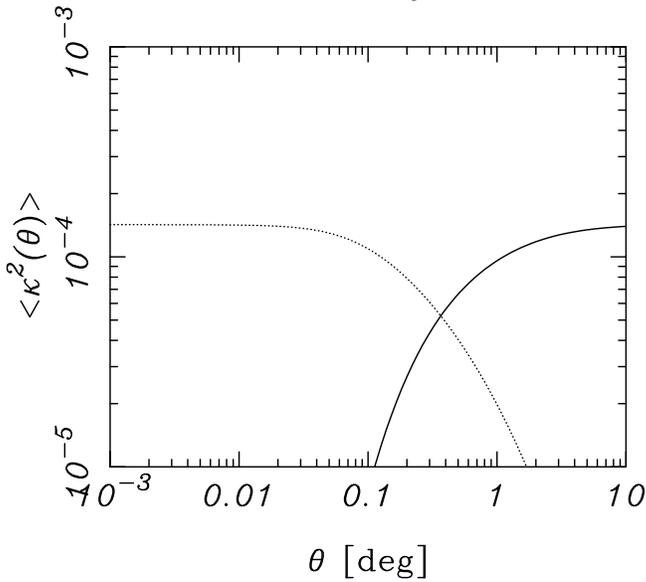}
\end{picture}
\caption{Variance of reconstructed convergence field, $\kappa'$,
Gaussian smoothed on a scale $\theta_s=0.05^\circ$, measured at
centre of Gaussian survey, as a function of survey radius,
$\theta$ (solid line). The variance in the reconstructed
convergence due to shear structure beyond the survey boundary is
also shown (dotted line).
 } \label{sampling_var_kappa}
\end{figure}

Figure \ref{sampling_var_kappa} shows the variance of the observed
convergence field, $\kappa'$, on a smoothing scale of
$\theta_s=0.05^\circ$, measured at the centre of a Gaussian survey
as a function of the survey radius, $\theta$ (solid line). We have
assumed a convergence power spectrum, $C^{\kappa \kappa}_\ell$,
for a flat LCDM universe with $\Omega_m=0.3$, $\Omega_\Lambda=0.7$
and used the Peacock-Dodds transformation (Peacock \& Dodds 1996)
to map to the nonlinear regime. Below a survey radius of
$\theta=0.1$ deg., missing modes due to the finite window and
mode-mixing result in a drop in the measured variance.

\subsubsection{Uncertainty in the reconstructed convergence}

In addition to estimating the variance in the reconstructed
convergence field $\lgl \kappa'(\thetab) \rgl_{\rm obs}$ measured from
within a finite survey, we can also predict the uncertainty $\lgl
\kappa'(\thetab) \rgl_{\rm miss}$ due to missing structure in the
shear field beyond the survey area. As the total variance measured
within the survey and the missing modes from beyond the survey must
yield the total variance of the convergence field $\lgl
\kappa'(\thetab) \rgl_{\rm total}$, the uncertainty in a
reconstruction due to missing structure is
 \be
        \lgl \kappa'(\thetab) \rgl_{\rm miss} =
            \lgl \kappa (\thetab) \rgl_{\rm total} - \lgl
            \kappa'(\thetab) \rgl_{\rm obs}.
 \ee
This uncertainty is plotted in Figure \ref{sampling_var_kappa}
(dotted line). For small survey radii, the variance in the
reconstruction uncertainty is just the total variance of the
convergence field, as the reconstruction uncertainty is dominated
by missing structure beyond the survey boundary. As the survey
radius approaches $\theta=0.1^\circ$, this effect begins to
decrease, and beyond 1 degree the survey is large enough to
include all the relevant structure.

\subsubsection{Variance of the differential lensing potential}

\begin{figure}
\centering
\begin{picture}(200,200)
\includegraphics{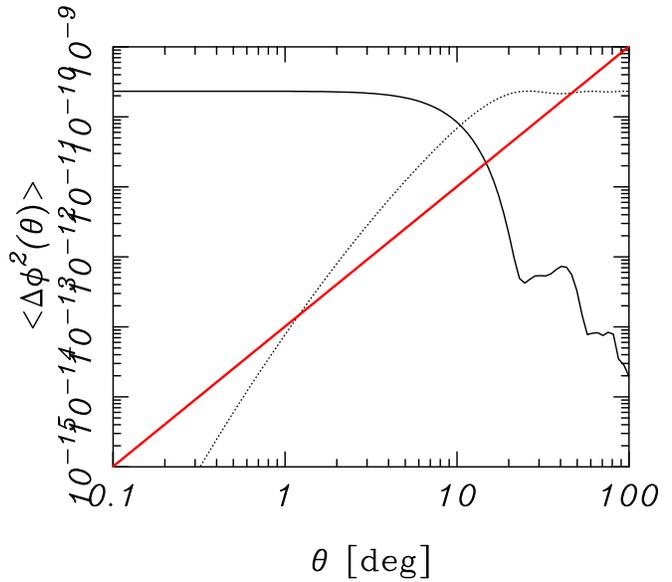}
\end{picture}
\caption{The effect of a finite window on the lensing potential
difference $\Delta \phi$ for a Gaussian window of radius $\theta$.
The dotted line is the sampling variance including mode-mixing
from the interior of a finite survey, while the solid line is
variance due to missing modes exterior to the survey. Oscillations
are real and due to the window function. The thick grey line is
the shot-noise estimate for a survey with $n_2=30$ galaxies per
sq. arcmin.} \label{sampling_var_phi}
\end{figure}

We can also calculate the variance of the differential lensing
potential field at the centre of a circular survey
 \be
    \lgl |\Delta \phi(0)|^2 \rgl = 4 \int\! \frac{d^2 \! \ell}{(2 \pi)^2}
        \ell'^{-4} C_{\ell}^{\kappa\kappa}
        |\widetilde{W}_{\ell}(0)|^2
            \left[ 1- \frac{2 J_1(\ell \theta)}{(\ell
            \theta)}\right]^2,
 \label{eq:dp}
 \ee
 where $J_1(x)$ is a Bessel function. The term in the square
 brackets subtracts off the mean field estimated over the survey
 area. We plot this  variance as a function of survey size, $\theta$,
 in Figure \ref{sampling_var_phi} (dotted line). We again assume a
 convergence power spectrum, $C^{\kappa \kappa}_\ell$, for a flat
 LCDM universe, and assumed the background galaxies are at $z=1.$

For small surveys the fluctuations expected in $\Delta \phi$ are
greatly reduced, but on large scales the expected variation on $\Delta
\phi$ becomes large; $\Delta \phi \approx 10^{-5}$. This is due to the
$\ell^{-4}$ weighting factor in equation (\ref{eq:dp}) which makes the
$\Delta \phi$ field sensitive to very large-scale structures. This is
just a reflection of the long-range nature of the 2-D lensing
potential field, but means that the peak variance is on large angular
scales. For smaller surveys these differential variations are
suppressed as the survey becomes smaller than the structure causing
them.

We conclude from this that the lensing potential field requires a
large survey for a complete sampling of modes.  For a LCDM model
we find that the potential modes are only fully sampled for
surveys above $10$ degrees. On the other hand, when wishing to
reconstruct cluster-scale mass concentrations, it is helpful to
restrict a reconstruction to a cell-size of $\sim 1$ sq. deg. as
this will cut out the large $\phi$ fluctuations due to
larger-scale structures (c.f. the good signal-to-noise obtained in
this fashion for cluster reconstructions in Sections 5 and 6).

\subsubsection{Uncertainty in the differential lensing potential}

As well as the intrinsic variance of the differential 2-D lensing
potential which we measure in a field, $\lgl |\Delta \phi(0)|^2
\rgl_{\rm obs}$, we can also calculate the uncertainty in lensing
potential due to missing modes from beyond the survey scale, $\lgl
|\Delta \phi(0)|^2 \rgl_{\rm miss}$:
 \be
         \lgl |\Delta \phi(0)|^2 \rgl_{\rm miss} = \lgl |\Delta \phi(0)|^2
         \rgl_{\rm total}-  \lgl |\Delta \phi(0)|^2 \rgl_{\rm obs}.
 \ee
 This is also plotted in Figure \ref{sampling_var_phi} (solid
 line). For small survey radii we again see that the uncertainty
 in the reconstruction is dominated by the missing shear structures
 beyond the survey boundary. In this case most of this missing
 structure is on larger scales. As we reach a survey radius of around 10
 deg. the survey begins to include this important large-scale
 shear structure, and the reconstruction uncertainty drops.

Also plotted on Fig \ref{sampling_var_phi} is the shot-noise
contribution to the uncertainty on a reconstruction from equation
(\ref{phi_err}), assuming $n_2 = 30$ galaxies per sq. arcmin (thick
solid line), typical for a ground-based survey. We see that shot noise
is larger than the expected rms $\phi$ fluctuations arising from
large-scale structure for very small surveys ($\theta < 1$ degree). On
larger scales the shot noise is lower than the expected signal,
allowing mapping of large-scale structure with good signal-to-noise on
these scales.

We note that the incompleteness contribution dominates over
the shot-noise contribution for small surveys, while the shot-noise
contribution dominates for large surveys. The two contributions are
around the same magnitude at around $\theta=15^\circ$ when $\lgl
\Delta \phi ^2\rgl \approx 2 \times 10^{-11}$.

\subsection{Wiener filtering lensing fields}

As we shall see, realistic galaxy ellipticities create a shot
noise contribution to the gravitational potential reconstruction
which far exceeds the expected gravitational potential amplitude
from a cluster. This suggests one of two approaches to
reconstructing gravitational potentials in practice: we can either
examine the potential statistically from many objects of interest
(e.g. stacking the signal from many groups or clusters); or we can
filter the signal to overcome the large noise contribution. A
valuable approach which we will use later involves Wiener
filtering the gravitational potential (c.f. Hu \& Keeton 2002).

In order to apply this filtering, we constuct the vector $\mathbf{P}$
containing our gravitational potential measurements along a particular
line of sight. We also calculate $\mathbf{N}$, a matrix containing the
noise covariance of the gravitational potential radially along this
line of sight; this can be measured directly from many $\Phi$
reconstructions of a zero $\phi$ field with appropriate noise. Finally
we require a matrix $\mathbf{S}$, representing the expected covariance
of the real gravitational potential signal along the line of sight. We
use a multiple of the unit matrix for $\mathbf{S}$, with amplitude
chosen to equal the square of the expected gravitational potential
amplitude, e.g. for clusters. Then we can apply a Wiener filtering
\begin{equation}
\label{eqn:wiener}
\mathbf{R} = \mathbf{S} (\mathbf{N}+\mathbf{S})^{-1} \mathbf{P}
\end{equation}
where $\mathbf{R}$ is our desired filtered gravitational
potential. This filter uses our knowledge of the noise amplitude
and covariance, together with the expected signal amplitude, to
significantly reduce the impact of the noise. We will test the
practicality  of this approach in Sections 5 and 6.

\subsection{Photometric redshift errors}

In the above analysis, we assume that the distances to galaxies have
been estimated from redshifts, measured either spectroscopically or
photometrically. However, these redshifts include contributions from
local velocities as well as the velocity of the overall Hubble
flow. Thus the redshifts will cause a somewhat biased scatter in our distance
measurements, as the local velocities are generated from gravitational
instability due to the local Newtonian potential, and therefore
correlate with our mass estimates. We must assess the level of error
associated with this effect.

We first consider the effect of a random distance error, which is
significant for photometric redshifts. We assume the source positions
are perturbed by $\r \rightarrow \r + \epsilon(\r) \rhat$, where
$\epsilon(r)$ is a random field with zero mean and correlations $\lgl
\epsilon(\r) \epsilon (\r') \rgl = \sigma_\epsilon^2
\delta_D(\r-\r')$. Expanding the lensing potential we find that the
observed Newtonian potential field becomes
\begin{equation}
\phi'(\r) = \phi(\r) + \epsilon \de_r \phi(\r) .
\end{equation}
This contributes to the first-order uncertainty in the Newtonian
potential
\begin{equation}
\Delta \Phi = \sigma_\epsilon \left(\de_r \Phi - \frac{2}{r}
\left[ \Phi - \frac{1}{r} \int^r_0 \! dr' \, \Phi(\r') \right]
\right).
\end{equation}
At large distances from the observer the terms in the square brackets
vanish, and the leading contribution to the uncertainty in distance
comes from the gradient of the Newtonian potential. Since
$\sigma_\epsilon$ is, for appropriate redshift surveys, small in
comparison with the redshift depth probed, and since $\de_r \Phi$ will
be small in a smoothed survey, this effect should not be dominant in
recovering the gravitational potential. To confirm this, we will
examine the effect of redshift errors on our simulations in Section
6.1.

\subsection{Redshift-space distortions}

We must now consider the effect of velocity distortions, which may be
significant for the more accurate spectroscopic redshifts. In linear
theory, the velocity field is related to perturbations in the
mass-density field by

\begin{equation}
\vb = - H f(\Omega_m) \nablab \nabla^{-2} \delta
\end{equation}
where $f(\Omega_m) = d \ln \delta / d \ln a \approx \Omega_m^{0.6}$ is
the growth index of density perturbations (e.g. Peebles 1980). The
position of galaxies are then shifted into redshift space by

\begin{equation}
\r \rightarrow \s = \r + u(\r) \rhatb
\end{equation}
where $\s$ is the redshifted position and $u(\r)=\rhatb . \vb(\r)$ is
the radial component of the velocity field. The distorted
redshift-space lensing potential is then

\begin{equation}
\phi^s(\s) = \phi(\r) - \left(\frac{2 a \lambda_H^2 f}{3
\Omega_m}\right) \de_r \Phi \de_r \phi(r) .
\end{equation}
In this case the systematic distortion of the Newtonian potential is
second-order:

\begin{equation}
\Delta \Phi = - \left( \frac{2 a \lambda_H^2}{3 \Omega_m^{0.4}}
\right) [\kappa \de_r^3 \Phi + 2 \Phi \de_r^2 \Phi + (\de_r
\Phi)^2 ].
\end{equation}
The magnitude of this effect on the reconstructed density field is
$\Delta \delta \approx f \delta^2$, and so only contributes to second
order.

\subsection{Multiple scatterings}

A further concern is the fact that a fraction of light rays will be
multiply scattered as they travel from source to observer. How will
this affect our assumption that the shear field can be derived from a
lensing potential?

The scattering of light rays can be written as

\begin{equation}
\delta \theta_i' = D_{ij}(\thetab) \delta \theta_j
\end{equation}
where $D_{ij}$ is the lens distortion matrix, defined for a single
scattering by

\begin{equation}
D_{ij} = \delta^K_{ij} + \de_i \de_j \phi = (1-\kappa)
\delta^K_{ij} + \gamma_{ij}.
\end{equation}
The distortion matrix for $n$ multiple scatterings is then just the
product of distortion matrices,

\begin{equation}
D_{ij}^n = D_i^{(1) \, k_1} D_{k_1}^{(2)\, k_2} \cdots D_{k_n
j}^{(n)}
\end{equation}
where $D^{(i)}$ is the effect of scattering off the $i^{th}$ structure
along the light path.

For the case of double scattering we can then define an effective
convergence, shear and a rotation,

\begin{eqnarray}
\kappa_{\rm eff} &=& \kappa_1 + \kappa_2 - 2 \kappa_1 \kappa_2 -
\Tr \, \gamma_{1\, i}^k \gamma_{2 \, kj} \nn \gamma^{\rm eff}_{ij}
&=& \gamma_{1 \, ij}+\gamma_{2 \,ij} - \kappa_1 \gamma_{2 \, ij }
- \kappa_2 \gamma_{1\, ij} \nn & & + \gamma^k_{1\, (i} \gamma_{2
\, j)k} - [\Tr \, \gamma_{1\, i}^k \gamma_{2 \, k}^j]
\delta^K_{ij} \nn \omega_{ij} &=&\gamma_{1\, [i}^k \gamma_{2 \,
kj]}.
\end{eqnarray}
Thus, even for a double scattering, the presence of a rotational
component to the distortion matrix shows that the distortion can no
longer be strictly constructed from a potential. However, in the case
of weak lensing at the $10\%$ level, we see that the rotational
components will on average be around $0.1\%$. However we can expect
this to break down in the strong lensing regime, very close to massive
cluster centres.  Therefore it is worth considering how common projections
may be in lensing.

If we assume that the cluster spatial distribution is random, the
probability of finding one or more clusters at random along a given
line of sight of volume $V= 4 \pi f R^3/3$ is
\be
    P(N \ge 1) =  1- e^{-\lambda} \approx \lambda  \approx N_c f,
\ee
where $N_c$ is the number of clusters in the whole sky volume and $f$ is
the angular fraction of the sky covered by our line of sight. The
probability of seeing two or more clusters along a given line of sight is
\be
    P(N \ge 2) = \frac{1}{2} \lambda^2
\ee
The statistic we require is the conditional probablity of finding a
second cluster when we have found one,
\be
    P(2|1) = \frac{P(N \ge 2)}{P(N \ge 1)} \approx \frac{1}{2}
    \lambda.
\ee
This yields
\be
    P(2|1) \approx 3.4 \left(\frac{N_c}{10^5}\right)
        \left(\frac{\theta_{\rm cl}}{10'} \right)^2 \%,
\ee
where $\theta_{\rm cl}$ is the angular size of a typical cluster.
This calculation is highly approximate, but allows us to see that
finding clusters behind clusters can occur with non-negligible
probability. This will cause no difficulties for our method, except in
the strong lensing regime very close to the centre of the
clusters. Indeed, our method is a useful means of measuring several
mass concentrations along the same line of sight.

\section{Simulating 3-D Lensing}

In order to investigate the practical application of the reconstruction
formalism, we have conducted a series of simulations representing
a realistic space volume for a lensing survey, including galaxies
with an appropriate spatial distribution and intrinsic ellipticity
and gravitational lenses of appropriate size and mass. The shear
and magnification for objects behind the lenses can be calculated
(retaining the full information regarding 3-D variation of these
quantities), and the objects' shapes can be altered accordingly.
With this flexible simulation package in place we can attempt to
reconstruct the gravitational potential causing the lensing, using
knowledge of only the galaxy ellipticities, their redshifts, and
equations (\ref{inv}) and (\ref{kaiser-squires}). Here we describe
in detail the form of these simulations.

Immediately we are faced with a question as to which of the fields
described so far $(\gamma, \kappa, \phi, \Phi, \delta)$ we should use
for 3-dimensional analysis. In reality, the most appropriate field to
use depends upon the application intended. For detection and
measurement of mass concentrations along the line of sight (Section
7), the fields $\kappa$ and $\phi$ are the most useful, as they have
the best signal-to-noise ($\simeq 7\sigma$ in each case in [9
arcmin$^2$, $\Delta z = 0.05$] pixels, for field radius $0.5^\circ$;
see Section 6) and are transversely local representations of the mass
present. For a direct mapping of the gravitational field, $\Phi$ is
most appropriate for mapping particular concentrations such as groups
and clusters, as it has much less noise than the $\delta$ field on
$<1^\circ$ scales (c.f. equation (\ref{delerr2})). Indeed, as the
noise in the $\delta$ field grows quadratically with survey size
(equation (\ref{delfinerr})), as does the noise in the $\Phi$ field,
it may be that $\Phi$ is most useful at large survey areas as well,
depending on the required application. In this paper we concentrate on
reconstructions of cluster size mass concentrations, and will
therefore make use of the $\phi$ and $\Phi$ fields.

\subsection{Constructing the Shear Field}

\begin{figure}
\psfig{figure=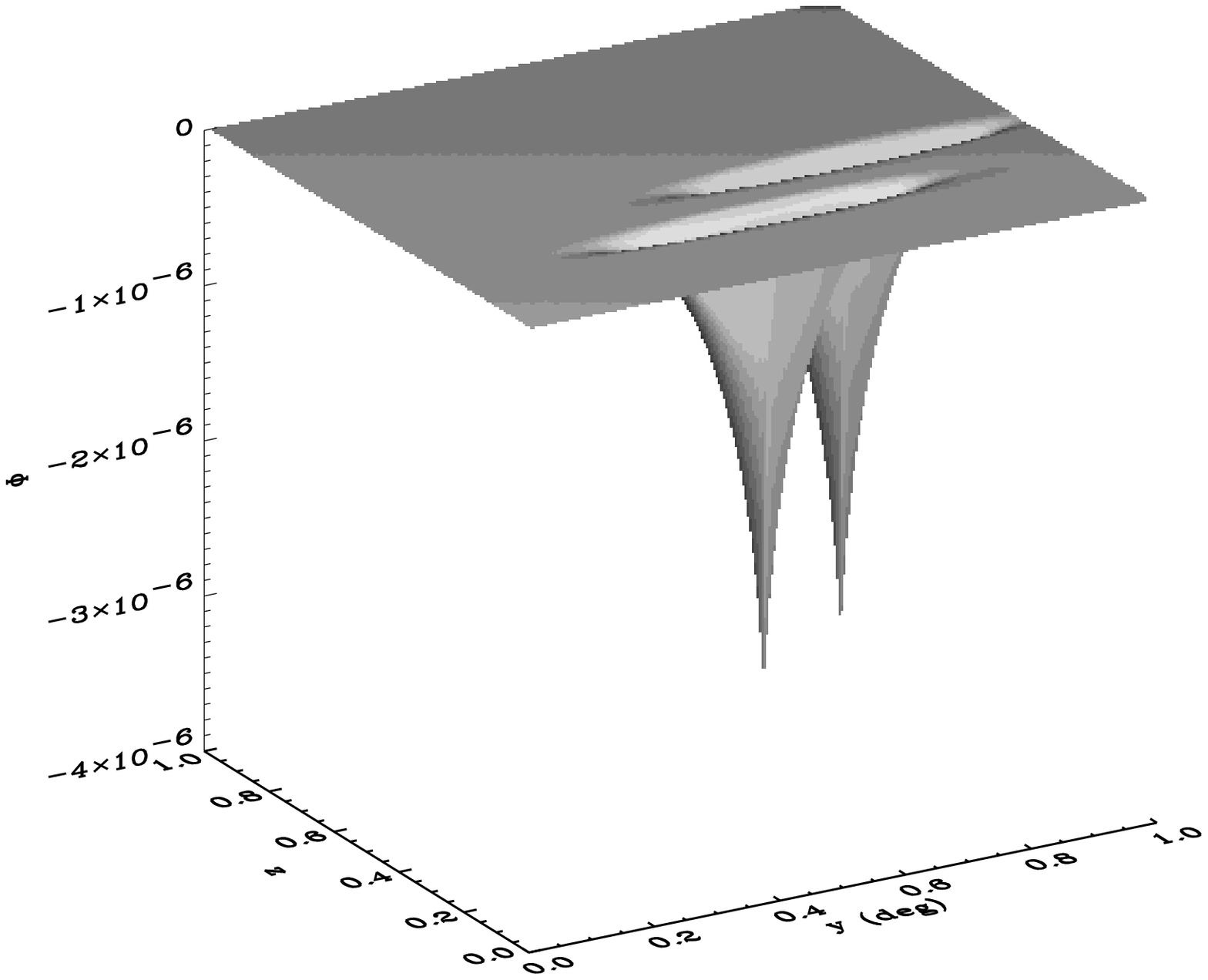,width=80mm}
\psfig{figure=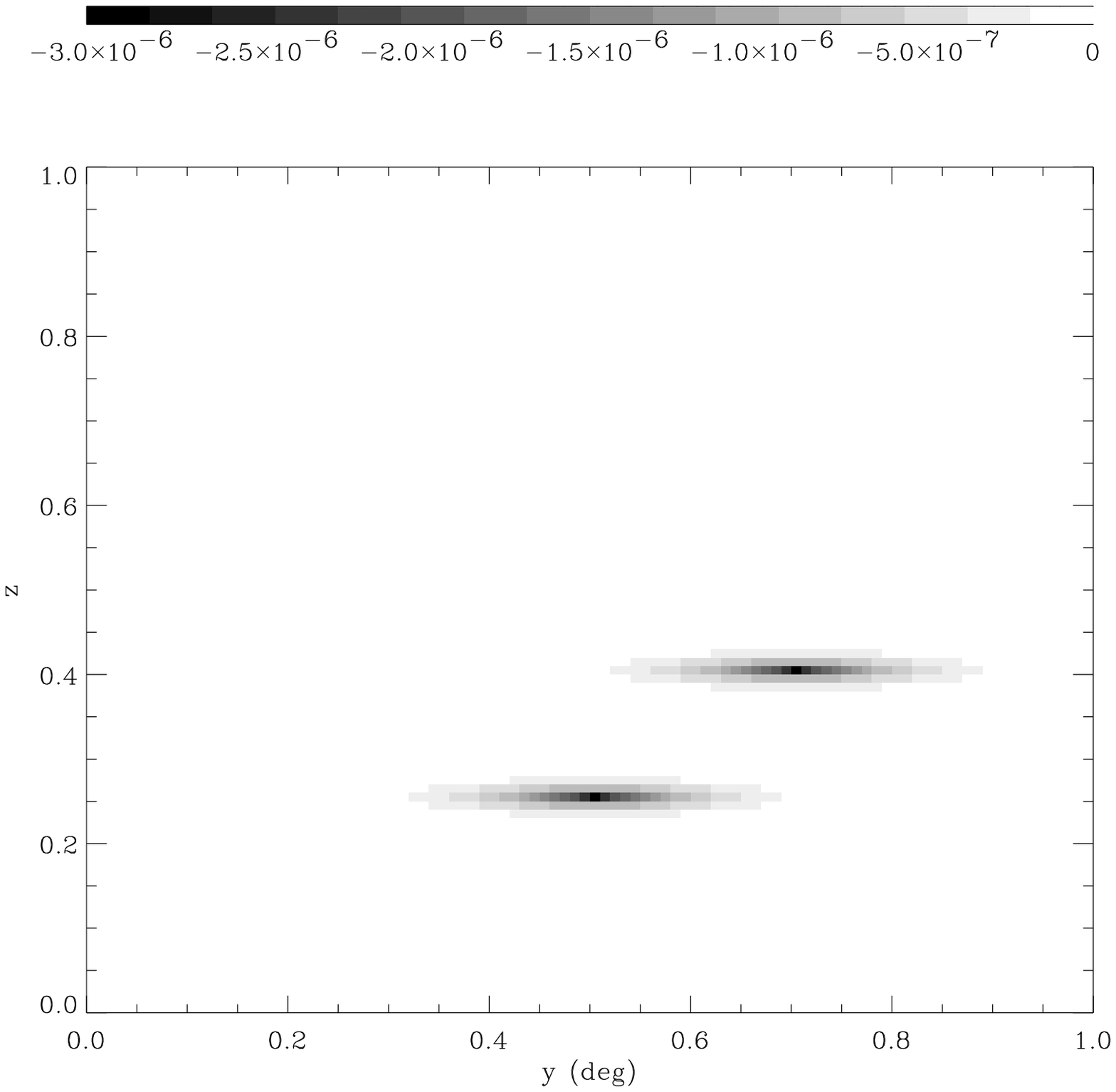,width=80mm} \caption{Example of
gravitational potential. Here we simulate the gravitational
potential for two NFW clusters at ${\mathbf r}=(0.5,0.5,0.25)$ and
$(0.5,0.7,0.4)$; the upper panel is a 3-D representation of
($y,z,\Phi$) resulting from an $x$ slice at $x=0.5^\circ$; the
lower panel displays the $\Phi$ values for ($y,z$) in greyscale.}
\label{fig:grav1}
\end{figure}

We construct a 3-dimensional grid, typically with a total of $100^3$
points. This represents the redshift cone in which we will attempt to
reconstruct the gravitational potential, i.e. the $x$ and $y$
directions represent an angular range on the sky, while the $z$
direction represents a radial distance. Typically we will use this to
model angular scales in $x$ and $y$ of $1^\circ$ while probing in the
$z$ direction down to an effective redshift of 1. We will therefore
quote coordinates in $\r=(x,y,z)$ ranging from 0 to 1. This does not
imply that the $z$ coordinate represents redshift, however; throughout
this paper, $z$ represents a comoving distance measure.

This choice of coordinates significantly simplifies our analysis: in a
flat universe, the comoving transverse separation of unperturbed light
ray paths converging at an observer is proportional to the comoving
radial distance along the light path (see e.g. Bartelmann \& Schneider
2000, Section 6.1). Thus unperturbed rays will simply move along our
$z$ coordinate with fixed $(x,y)$. Also, in order to lay down e.g. a
constant 3-dimensional comoving number density of objects, we simply
allocate a constant number density in our coordinates; no correction
is necessary for a varying physical number density. Finally, the
unitless transverse derivatives required in equation (\ref{gij}) are
simply $\partial_x$ and $\partial_y$ in our coordinate system.

We fix in this grid the positions of lenses which we will wish to
recover. We assign to each lens a mass and three perpendicular scale
lengths for the size of its gravitational potential. In the following
section we will be concerned with typical galaxy cluster lenses, in
which case we assign masses in the range $0.5-5 \times 10^{14}
M_\odot$ and radii of $0.5-2$Mpc. We will position these clusters
between $z=0.25$ and $z=0.5$ while probing lensed galaxies down to
$z=1$, mimicking lensing studies of massive clusters (e.g. Tyson et al
1990, Kaiser \& Squires 1993, Bonnet et al 1994, Squires et al 1996,
Hoekstra et al 1998, Luppino \& Kaiser 1997, Gray et al 2002).

Given these masses and scale lengths, we can calculate the
gravitational potential over the entire 3-dimensional grid. We have
used Navarro, Frenk \& White (1996) profiles for the density,
\begin{equation}
\rho(\r) = q / |\r - \r_c| (1 + |\r - \r_c|)^2,
\end{equation}
where $q$ is a measure of the mass, and $\r_c$ are the coordinates of
the centre of the mass profile. Using Gauss's law and a further
integration and renormalisation, we find the resulting gravitational
potential for a mass concentration
\begin{equation}
\Phi(\r)= - \frac{q_1 r_s}{|\r|} \log \left( 1 + \frac{ |\r -
\r_c|}{r_s} \right),
\end{equation}
with $q_1$ as a measure of mass and $r_s$ the NFW scale parameter. The
total gravitational potential in the simulated volume of space is
taken to be the sum of the individual lens contributions.

Figure 4 shows an example of our constructed gravitational potential,
for two NFW profile clusters, placed at $z=0.25$ and $z=0.4$; the view
is of a 1 grid-unit slice at $x=0.5^\circ$. The apparent distortion of
the clusters is due to the scales chosen; we are examining a large
distance scale radially ($\sim 3000$Mpc), with much smaller distance
scales transversely ($\sim 10$Mpc). We will wish to reconstruct this
gravitational potential for all $(x,y,z)$ from shear information only.

The masses of the clusters in this example were chosen to mimic the
shear properties of clusters in Gray et al (2002), with mass $m =
8\times10^{13}M_\odot$ within a radius of 2 arcmin. Note that in order
to calculate $\phi$ and $\Phi$ we convert the $x,y$ coordinates from
degrees to radians, leading to the same $\phi$ and $\Phi$
normalisations as in Section 3.

From the gravitational potential shown in Figure 4, we calculate the
lensing potential $\phi$ given by equation (\ref{pot}). This is a
necessary step towards calculating the shear, which is all we will use
for potential reconstruction. For each point in $(x,y)$ we set
$\phi(x,y,0)=0$ and then use the discrete version of equation
(\ref{pot}),

\begin{equation}
\phi(x,y,z)= -2 \sum_{w=1}^z \frac{z-w}{z w} \Phi(x,y,w) \Delta w,
\end{equation}
where we must first use this equation to calculate $\phi(x,y,0.01)$
then $\phi(x,y,0.02)$, etc. In this fashion, we can calculate the
lensing potential for all points on our 3-dimensional grid.

Figure 5 shows the lensing potential calculated as above for the
example introduced in Figure 4. Note that the axes on this 3-D plot
are $(y,z,\phi)$; we are observing how the lensing signal grows with
depth. The figure shows the generic behaviour for all lensing; the
lensing potential due to distortion from a massive object becomes
stronger with increasing depth, but asymptotes to a finite value at
large $z$.

\begin{figure}
\psfig{figure=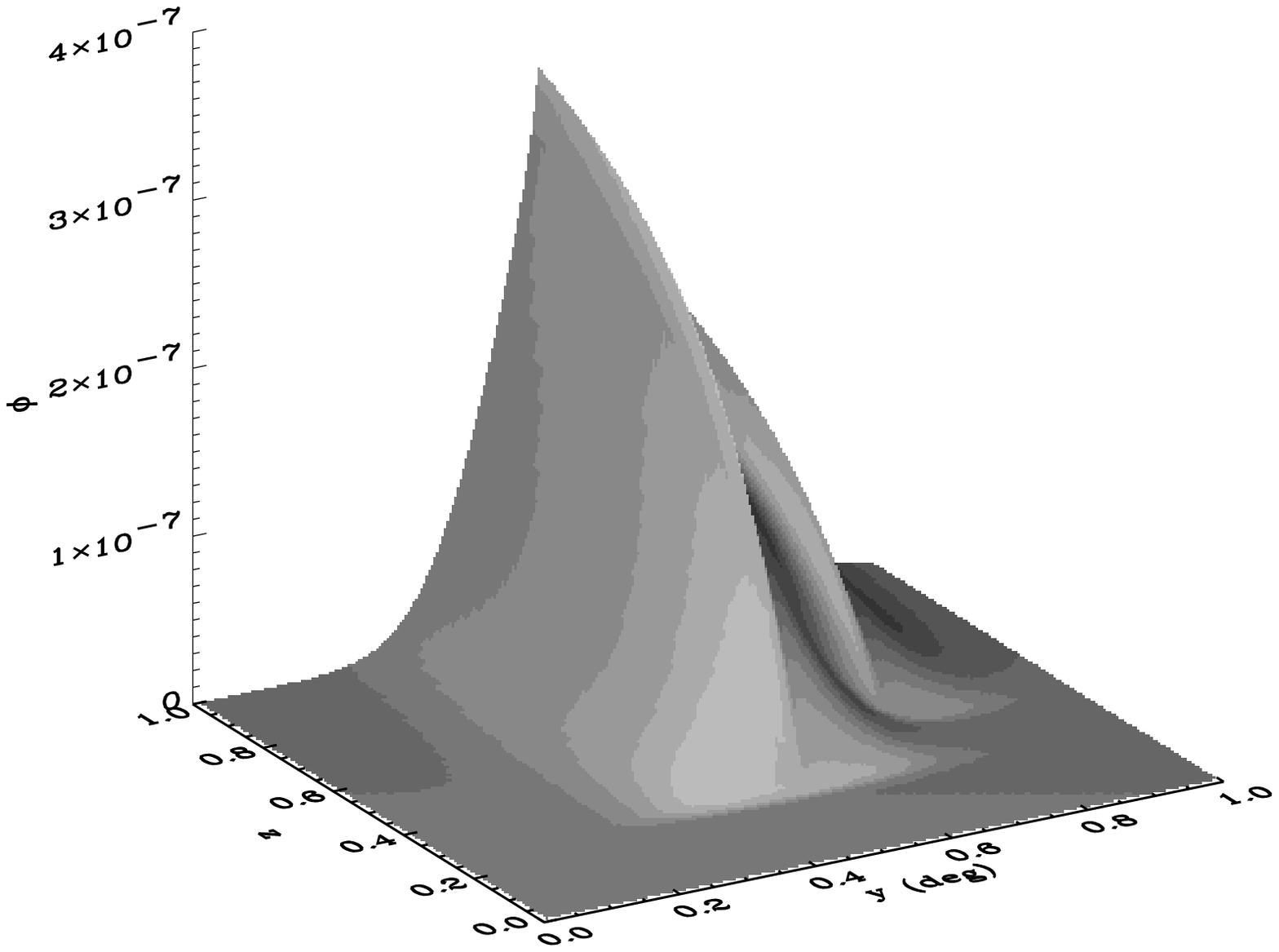,width=80mm}
\psfig{figure=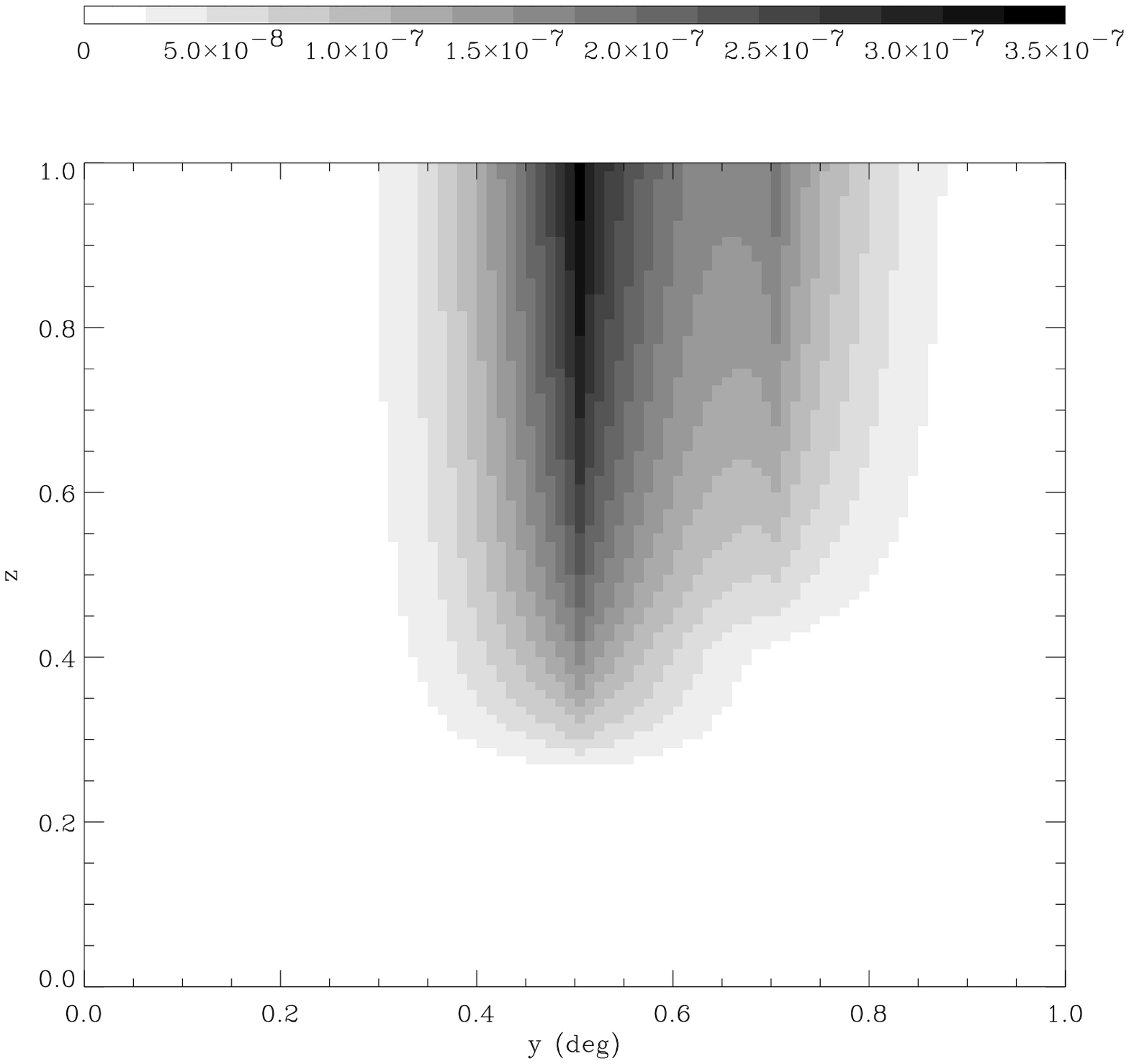,width=80mm}
\caption{Lensing potential calculated for the gravitational potential
of Figure 4. Note the increase in distortion expected with growing
$z$. In the upper panel, the axes represent the $(y,z,\phi)$
coordinates; in the lower panel the $\phi$ values for ($y,z$) are
shown in greyscale.}
\end{figure}

We are now in a position to calculate the 3-dimensional gravitational
shear field arising from the gravitational potential. We calculate the
shear components $\gamma_{1}$ and $\gamma_2$ from our 3-dimensional
$\phi$ field by first approximating
\begin{eqnarray}
\partial_{xx} \phi({\mathbf x})&\simeq&
[\phi(x+\Delta x,y,z)+\phi(x-\Delta x,y,z)\nonumber \\
& & -2\phi(x,y,z)]/(\Delta x)^2.
\label{dxxphi}
\end{eqnarray}
An entirely similar approximation is made for $\partial_{yy}
\phi({\mathbf x})$, while we approximate
\begin{eqnarray}
\partial_{xy} \phi({\mathbf x})&\simeq&\frac{1}{4} [\phi(x+\Delta
x,y+\Delta x,z)+\phi(x-\Delta x,y-\Delta x,z) \nonumber \\ & &
-\phi(x-\Delta x,y+\Delta x,z)\\&&-\phi(x+\Delta x,y-\Delta
x,z)]/(\Delta x)^2.
\label{dxyphi}
\end{eqnarray}
Then, following from equation (\ref{gij}), we can write the shear
components as

\begin{equation}
\gamma_1({\mathbf x})= -\frac{1}{2}(\partial_{xx}\phi({\mathbf
x})-\partial_{yy}\phi({\mathbf x}))
\end{equation}

\begin{equation}
\gamma_2({\mathbf x})= -\partial_{xy}\phi({\mathbf x})
\end{equation}
From these equations we can calculate shear values for all points on
our 3-dimensional grid. An example of a calculated shear field is
shown in Figure 6, corresponding to the gravitational potential of
Figure 4; this is an $(x,y)$ slice of the 3-D shear field at
$z=0.75$. Note the clear signatures due to the two clusters. A slice
further back in $z$ would have a similar $x,y$ pattern but a larger
shear magnitude; a slice in $z$ in front of the clusters would have no
shear signal.

\begin{figure}
\psfig{figure=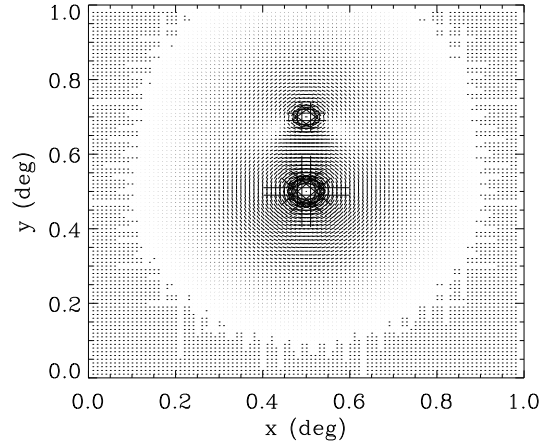,width=80mm}
\caption{Slice of the 3-D shear field corresponding to the
gravitational potential of Figure 4. This is an $(x,y)$ slice at
$z=0.75$. The largest $\gamma$ value plotted is 0.27.}
\end{figure}

We have therefore calculated the gravitational shear which would exist
for an object at any grid point; now we must place objects at some of
these grid positions, with an appropriate distribution in space. We
normalise the total number of objects to those expected in surveys
down to the required depth; for a ground-based survey probing to a
limiting redshift $z_r=1$, we expect a number density $n \simeq 30$
usable galaxies per square arcmin (e.g. Bacon et al 2002). Alternatively, we
can project the number densities expected for deep, space-based
surveys down to $z_r=1$ ($n \simeq 100$; see Massey et al 2002).

The selected total number density is used to find the probability of
an object existing at a given grid point in the following fashion. We
treat each $(x,y)$ sheet as a slice of a comoving pyramid (not a box),
in order to take into account the fact that the survey has an angular
extent. In some of the simulations we will assume that the
3-dimensional number density is constant, while in others we will
adopt $dN/dz \propto z^2 \exp(-a z^2)$; the distribution will be
stated in each case.

In the case of varying $N(z)$, the number of galaxies $N$ expected per
grid point can be found using

\begin{equation}
N(x,y,z)= \frac {d N}{dz} \frac {\Delta z}{(\theta)^2}
\label{nz}
\end{equation}
where $\theta$ is the transverse survey length (usually $1^\circ$ in our
simulations), $\Delta z$ is the radial grid increment, and $N_{tot} = n
\Omega$, where $\Omega$ is the solid angle extent of the survey. For
each $z$, we then generate $N(z) \theta^2$ random coordinates
in $x$ and $y$, and increment the number of objects at the nearest
grid point for each coordinate pair.

For each object, a random Gaussian-distributed shear value is chosen
to simulate the effect of intrinsic ellipticity, and is added in
quadrature to the gravitational shear value. For typical ground-based
surveys the scatter in shear estimators due to the intrinsic
ellipticity can be well modelled by Gaussians in $\gamma_1$ and
$\gamma_2$ with standard deviation 0.3 in each component (e.g. Bacon
et al 2002). For space-based survey simulations, we use the smaller
standard deviation of 0.2 in each shear component (see e.g. Rhodes et
al 2001).

We now have a set of galaxies sampling the 3-D shear field at a finite
set of points in space, with an additional shot noise contribution
from their intrinsic shapes. We must now attempt to estimate the
underlying purely gravitational 3-D shear field, and from that to
calculate the 3-D lensing potential $\phi$ from equation
(\ref{kaiser-squires}), and the corresponding 3-D gravitational
potential $\Phi$ from equation (\ref{inv}).

\subsection{Reconstructing the Potential}

We can overcome the shot noise from galaxy ellipticities by a
combination of binning many galaxies' shears in a cell, and smoothing
the shear field. In the specific examples given below, we find it
convenient to initially rebin our galaxy grid to pixels with 3 arcmin
diameter in the transverse direction and $\Delta z=0.05$ in the radial
direction; further smoothing can be carried out as necessary
later. We calculate the averaged shear field for the rebinned grid;
this averaged shear will then be the basis of our potential
reconstruction.

We can conveniently find the $\phi$ field corresponding to this
$\gamma$ field using Fourier tranforms of these two fields. Following
Kaiser and Squires (1993), we find from equation
(\ref{kaiser-squires}) an optimal estimate of the lensing potential,

\begin{equation}
\phi({\mathbf k}) = \frac {2 ( k_x^2 - k_y^2 ) \gamma_1
({\mathbf k}) + 4 k_x k_y \gamma_2 ({\mathbf k})}{k_x^2 + k_y^2},
\end{equation}
where $\phi({\mathbf k})$ etc are Fourier transformed quantities. This equation
is valid if the transverse differential operators of equation
(\ref{kaiser-squires}) are dimensionless; this is satisfied by our
model, where the $x$ and $y$ directions represent angles rather than
physical distances.

In this fashion we calculate $\phi({\mathbf k})$ using Fast
Fourier Transforms, and hence find $\phi$. Since noisy low-$k$
modes can introduce rather large variability in $\phi$ with $z$
(see Section 2), we further add a constant at the edge of our
(real-space) field to ensure that the mean of $\phi$ around the
edge of the field is zero (this is equivalent to equation
(\ref{unbiaspot}) above; see Section 3.1.2 and 3.2 for a
discussion of the level of error this causes). We can then smooth
the $\phi$ field in the $z$ direction to reduce the noise
amplitude, convolving the $\phi$ field with a radial Gaussian. We
can choose the width of this kernel in each circumstance as a
compromise between reducing noise and retaining spatial
resolution.

We then calculate the gravitational potential $\Phi$ using equation
(\ref{inv}). As in equations (\ref{dxxphi}) and (\ref{dxyphi}), we use
local differences in $\phi$ in the $z$ direction to approximate the
derivatives. If necessary, we can smooth or filter $\Phi$ itself to
further reduce noise levels.

\begin{figure}
\psfig{figure=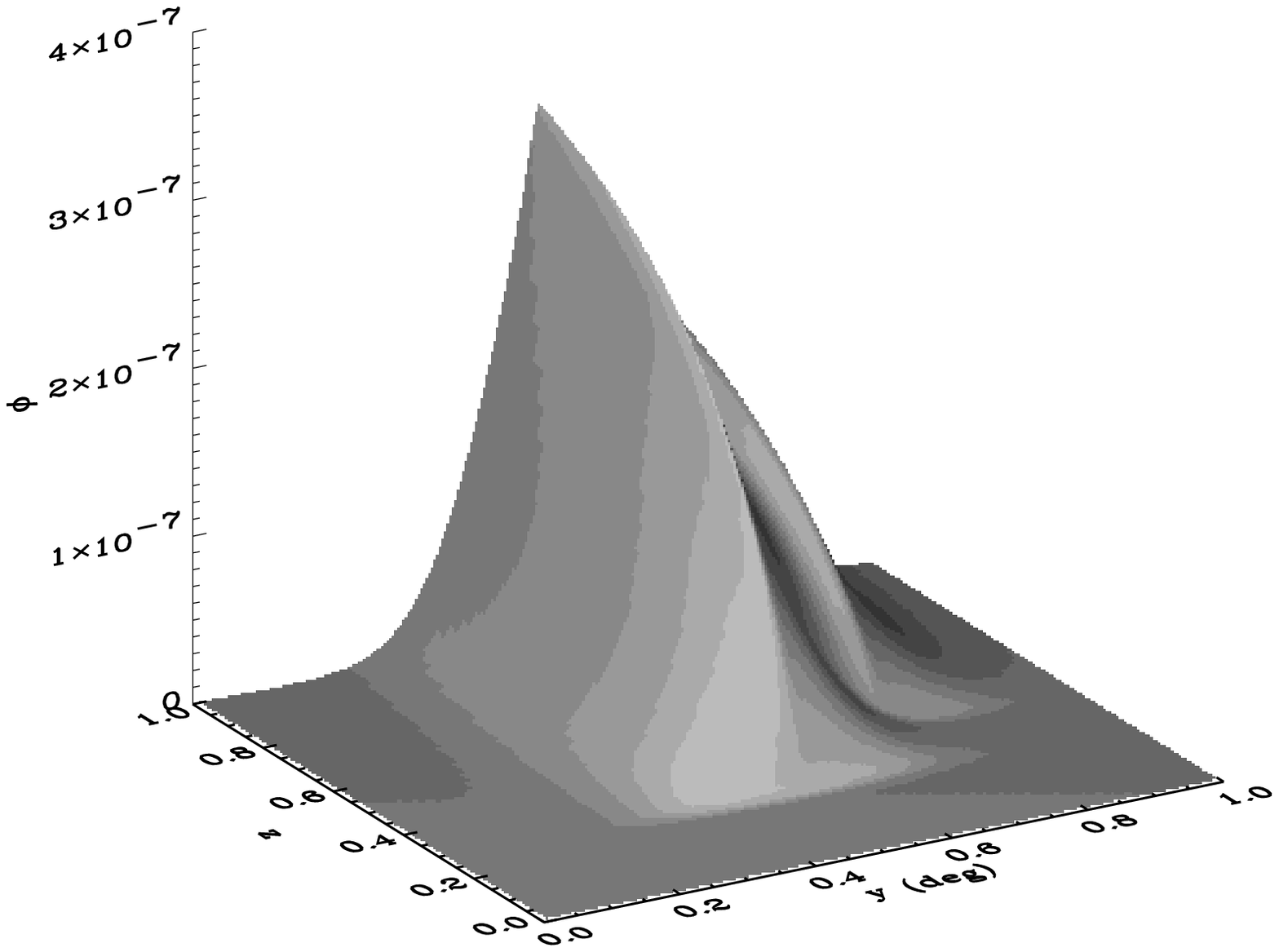,width=80mm}
\psfig{figure=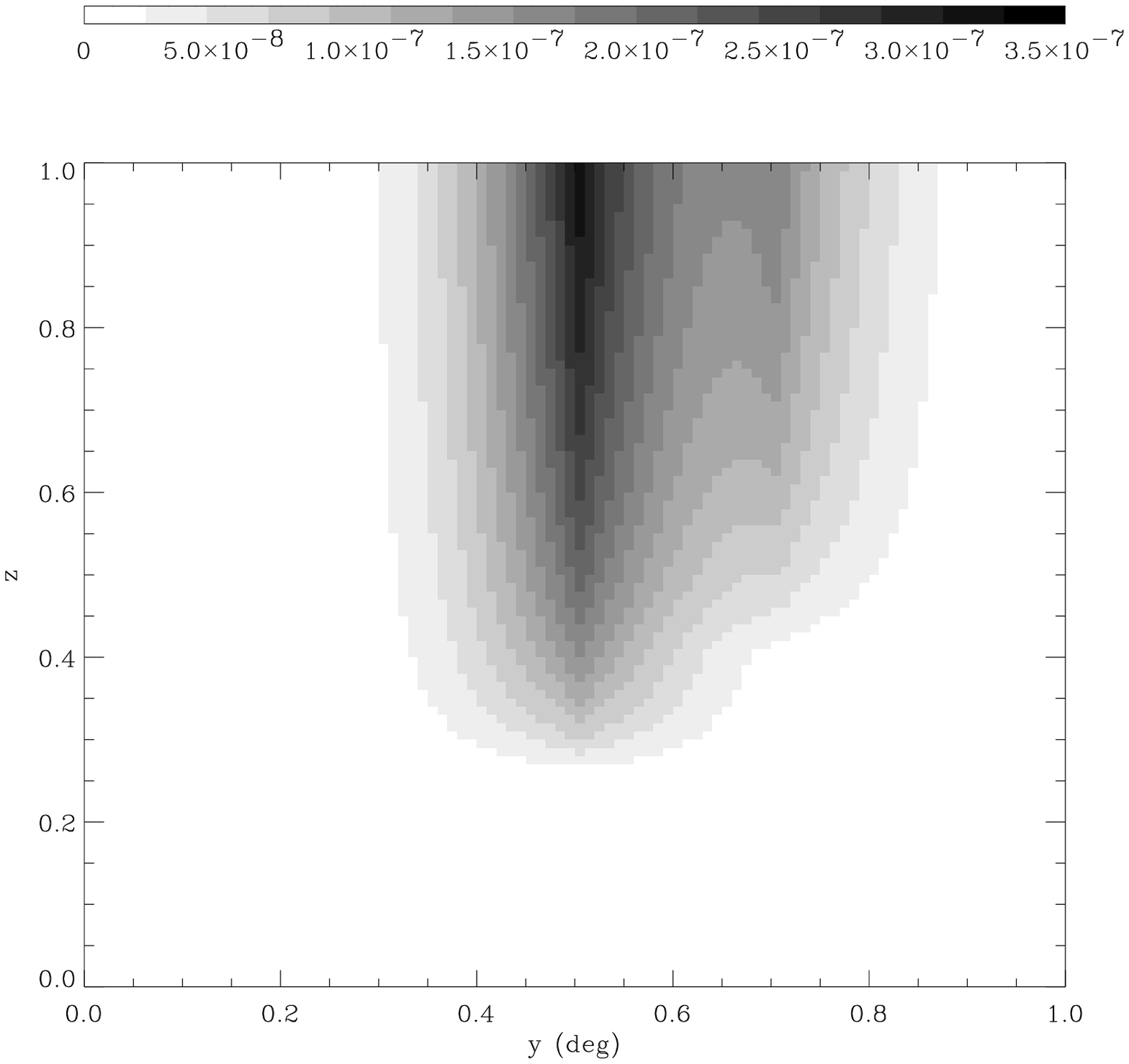,width=80mm}
\psfig{figure=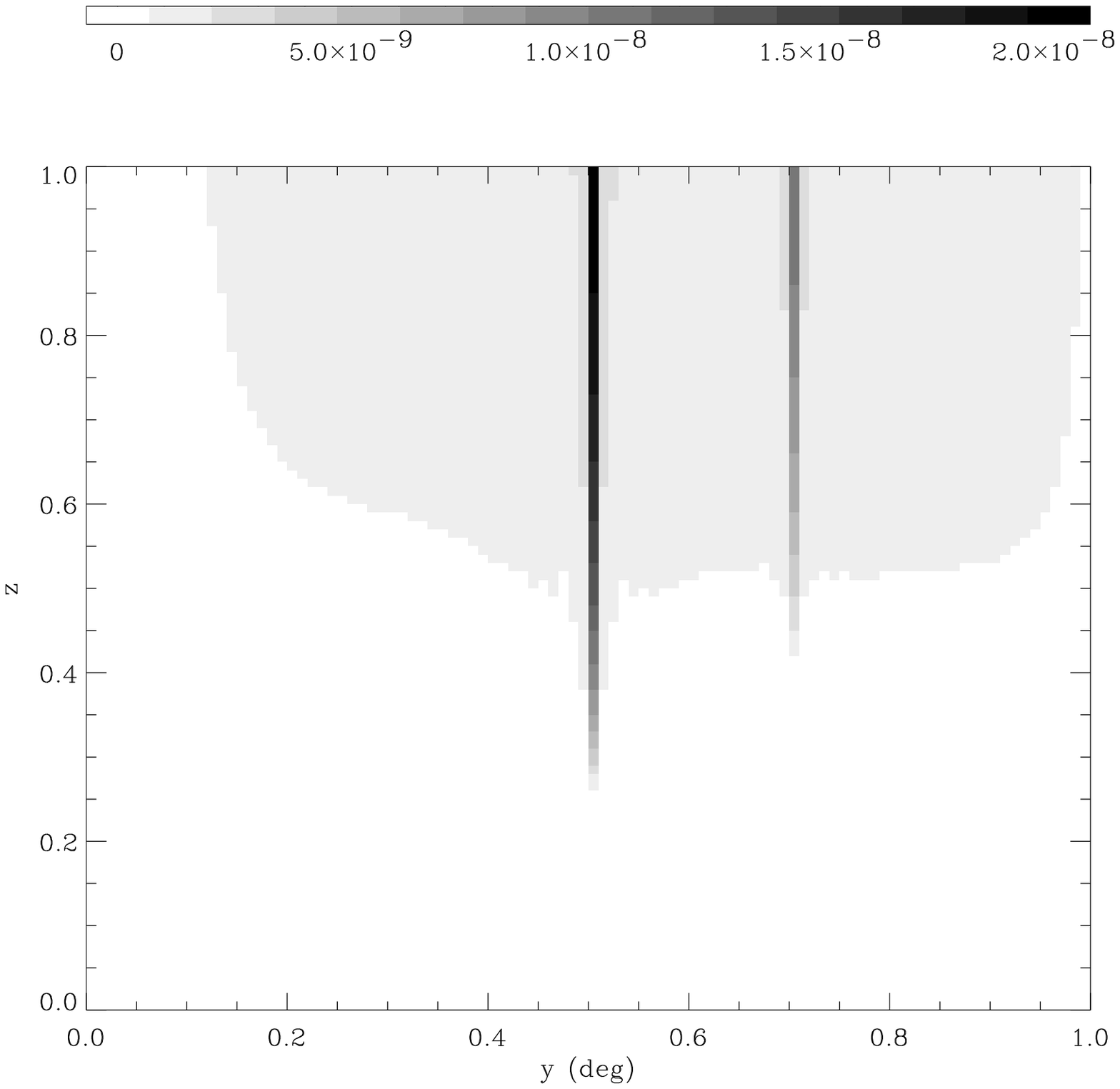,width=80mm}
\caption{Top and middle panel: reconstructed lensing potential using
the full shear field of Figure 6, in a $100^3$ grid. This is a $(y,z)$
plane at $x=0.5^\circ$. Bottom panel: difference between input and recovered
lensing potential fields.}
\end{figure}

\begin{figure}
\psfig{figure=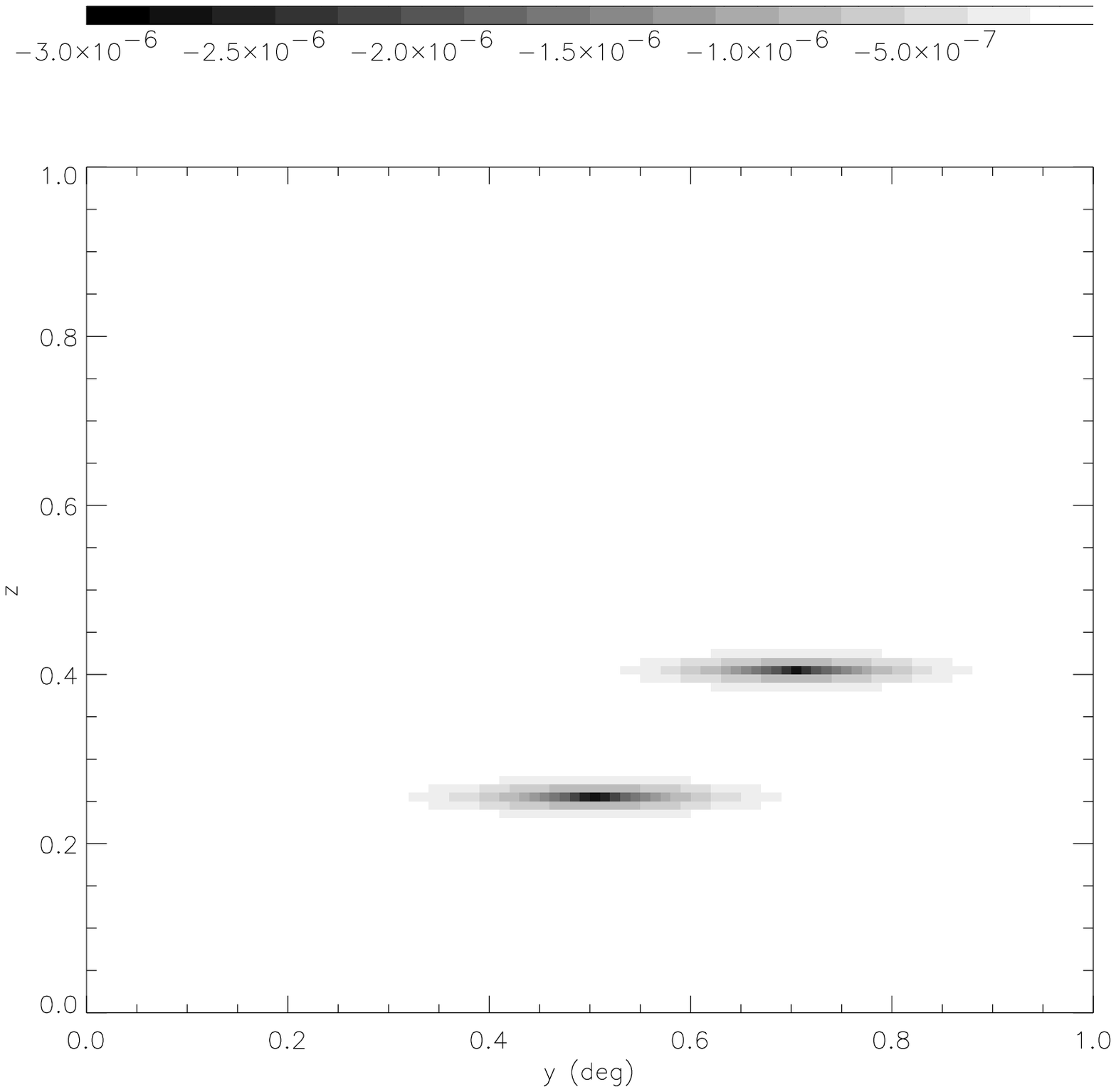,width=80mm}
\psfig{figure=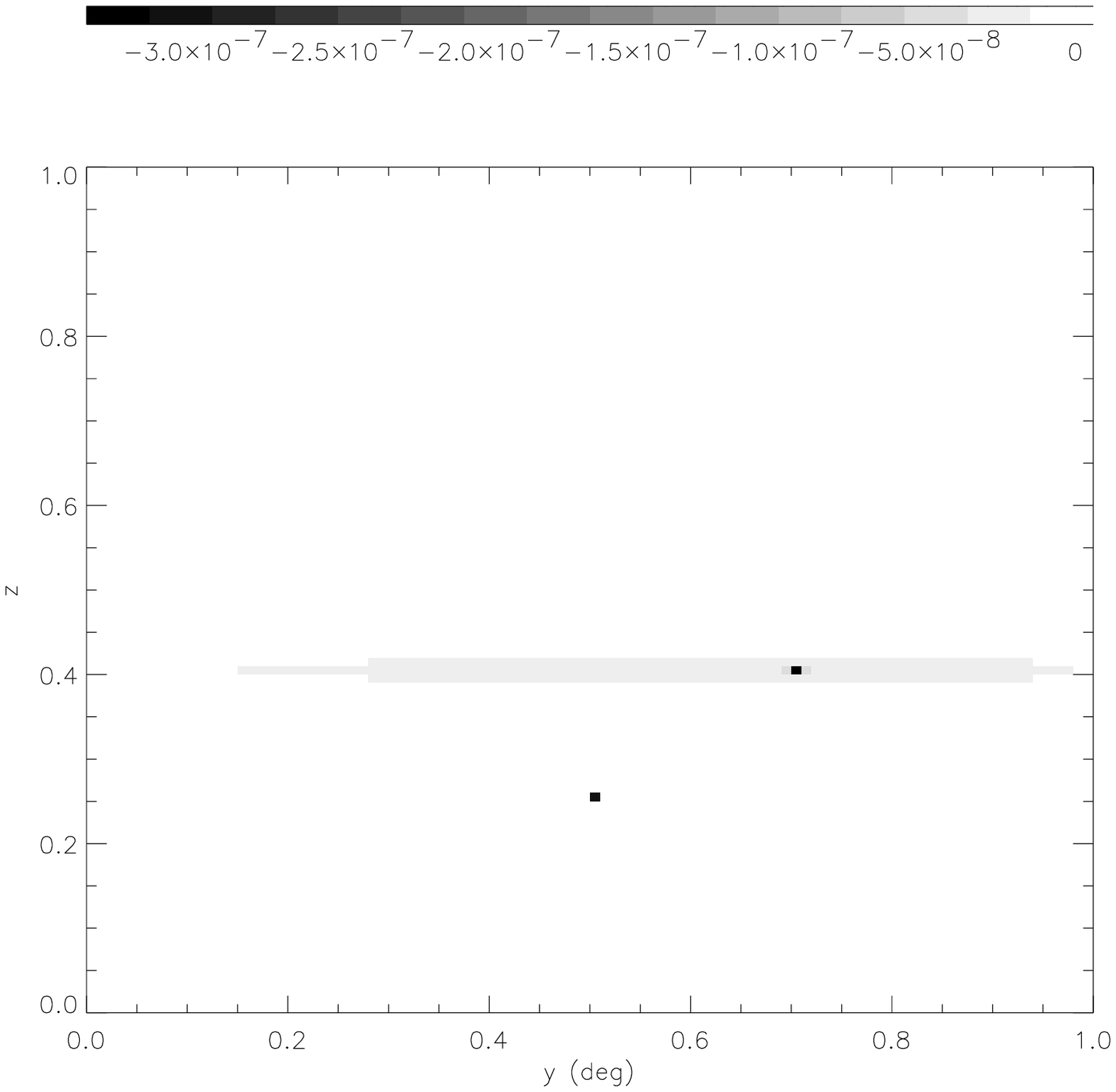,width=80mm}
\caption{Top panel: reconstructed gravitational potential using the
full shear field of Figure 6, in a $100^3$ grid. This is a
$(y,z)$ plane at $x=0.5^\circ$. Bottom panel: difference between input and
recovered gravitational potential fields.}
\end{figure}

\section{Simulation Results}

We have described above our means of simulating mass distributions and
corresponding shears, followed by our method for reconstructing the
lensing potential and gravitational potential given a finite number of
noisy estimators of the shear. Here we describe our findings for this
reconstruction.

\subsection{Perfect reconstruction}

First we examine the accuracy of our method when the shear field
is perfectly known everywhere. Figure 7 shows the reconstruction
of the lensing potential when we simply input into our
reconstruction the full shear field shown in Figure 6; we do not
smooth the shear field in this case. We regain the lensing
potential successfully; this is quantified in Figure 7c, where we
show the difference between the reconstructed and original $\phi$
fields, for a slice through the reconstruction. This error
introduced by our numerical implementation is $<3$\% of the
lensing signal everywhere behind the clusters within 0.15 degrees
of the cluster centres. The one exception is the core pixel line
behind each cluster, where the error is 6.6\% and 9.7\% of the
lensing potential for the left and right cluster respectively.
This is expected, due to the cusp within this pixel, which the
binned shear cannot accurately follow. We see that our procedure
for reconstructing the lensing potential is very successful in the
absence of noise.

We proceed to use our reconstructed lensing potential to find the 3-D
gravitational potential, still using the full shear field (i.e. still
without sampling by a finite number of objects, or adding shot noise
due to galaxy ellipticities). Our result is shown in Figure 8, which
displays the reconstructed $\Phi$ and the difference between input and
reconstructed fields. Again it can be seen that, with perfect
knowledge of the shear field, a good reconstruction is achieved with
our method. For the cluster at $z=0.25$, the error is $<0.4$\% of the
signal within a radius of 0.2 degrees of the cluster, except for the
core pixel where the error is 11.5\%. For the cluster at $z=0.5$, the
error is $<5\%$ of the signal within a radius of 0.1 degrees of the
cluster, except for the core pixel where the error is 17.6\%. This is
again due to the cusp at the cluster centre, which cannot be followed
well by our averaged shear field. Nevertheless, it is clear that our
method is successful in reconstructing cluster gravitational potentials
in the absence of noise.

\subsection{Reconstruction with noise}

Having demonstrated that the inversions of the shear field to obtain
lensing and gravitational potentials are viable in the absence of
noise, we now wish to add the two primary sources of noise present in
lensing experiments: Poisson noise due to only sampling the field at a
finite set of galaxy positions, and additional noise due to galaxies'
non-zero ellipticities. We use appropriate number densities for
plausible ground-based experiments (30 per sq arcmin) and space-based
experiments (100 per sq arcmin), and use equation (\ref{nz}) to place
an appropriate number of objects at each redshift slice. We continue
to use the gravitational potential of Figure 4.

We incorporate the effect of the intrinsic ellipticities of the
galaxies as described in Section 4 (i.e. adding a Gaussian-distributed
random shear value to the gravitational shear, with a standard
deviation of 0.2 per shear component for space-based applications and
0.3 for ground-based applications).

Given this noisy shear field, we carry out our reconstruction as
described in Section 4, with a Gaussian smoothing of the $\phi$ field
in the $z$ direction with a $1\sigma$ width of 0.1 in redshift (not
applied for calculating Wiener-filtered $\Phi$).

We will now discuss the reconstructions obtained for ground-based and
space-based data. In the discussion below, we will often describe measurement
significance in terms of $\nu({\mathbf x})=I({\mathbf
x})/\sigma({\mathbf x})$, with $I({\mathbf x})$ the amplitude of the
lensing potential or gravitational potential at a particular point
${\mathbf x}$, and $\sigma({\mathbf x})$ the noise level in the
vicinity of this point.

\subsubsection{Lensing Potential for Space-based Experiment}

\begin{figure}
\psfig{figure=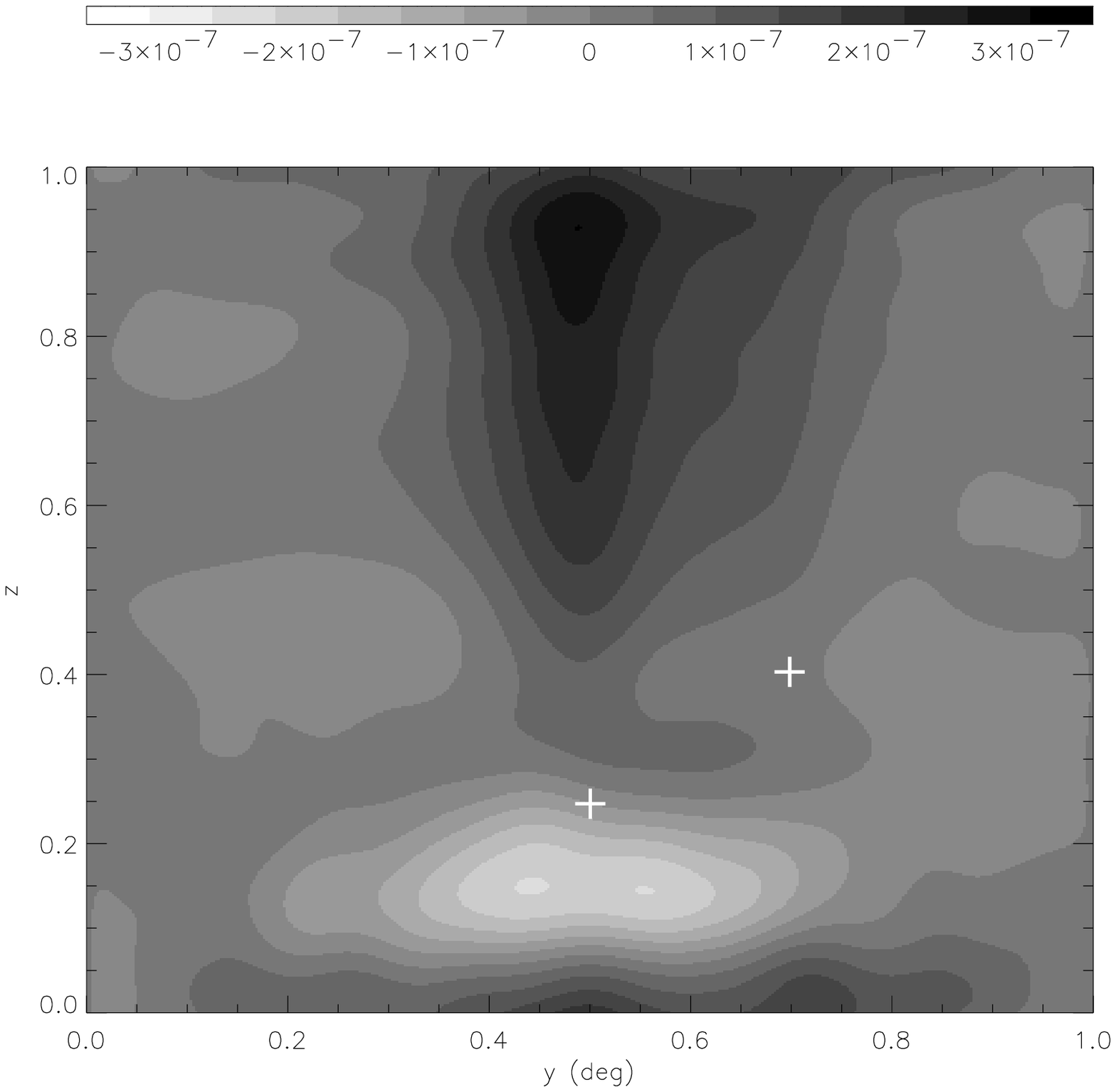,width=80mm}
\psfig{figure=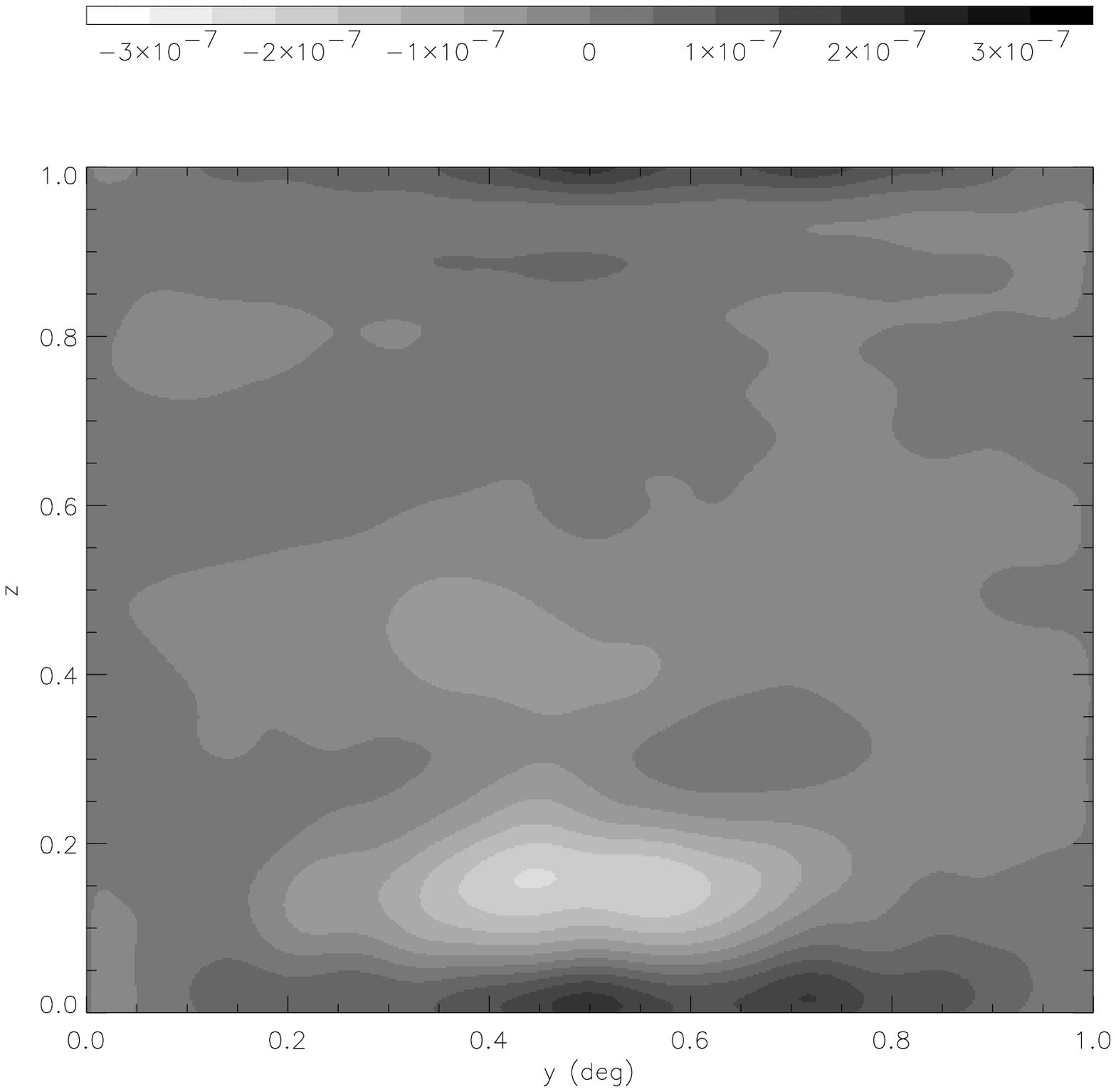,width=80mm}
\caption{Top panel: reconstructed lensing potential $\Delta \phi$
using finite number of galaxies with realistic ellipticities; $n=100$
per sq arcmin, $\gamma_{\rm rms} = 0.2$ per shear component, as
expected for a notional space-based survey. Crosses show the positions
of the cluster centres. Bottom panel: difference between input and
recovered lensing potential fields.}
\end{figure}

\begin{figure}
\psfig{figure=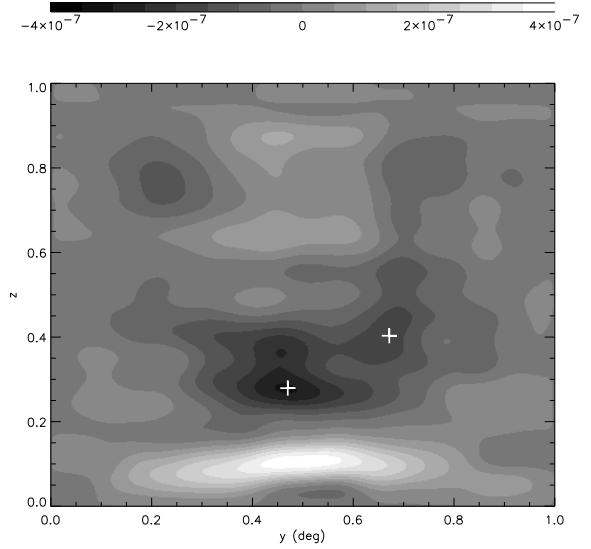,width=80mm}
\caption{Reconstructed gravitational potential for our space-based
experiment, using galaxy properties as in Figure 9, after Wiener
filtering. Note the detections of clusters at $(y,z)=(0.5,0.2)$ and
$(0.7,0.4)$ with $\nu=4.2$ and 2.1 respectively (crosses show trough
minima).}
\end{figure}

The resulting lensing potential reconstruction for our space-based
experiment is shown in the top panel of Figure 9. (In this and later
figures, we plot the results for our $20^3$ grid, but display a
resampled grid calculated by padding the Fourier transform of the grid
with high-$k$ modes set to zero.) We find that we obtain a reasonable
reconstruction of the lensing potential, with $\nu \simeq 6.9$ per
pixel in the background ($z > 0.75$) for the nearer cluster, with $\nu
\simeq 4.2$ for the $z=0.4$ cluster, within 0.1 deg radius of the
cluster centres. We could increase this signal by rebinning or
smoothing, at the cost of reducing spatial resolution. We could also
find an overall signal-to-noise for each cluster by finding a means of
averaging all of the lensing signal arising behind a cluster; we will
discuss this in Section 7. Note the large noise peaks in the
foreground ($z<0.2$) of the reconstruction, due to the small number of
objects available in this volume.

The lower panel of Figure 9 shows the difference between the input and
recovered lensing potential. Pleasingly, we observe no evidence of
residuals associated with misconstruction of the lensing
potential. The noise levels are as expected from equation
(\ref{phi_err}), as discussed in Section 6 below.

\subsubsection{Gravitational Potential for Space-based Experiment}

\begin{figure}
\psfig{figure=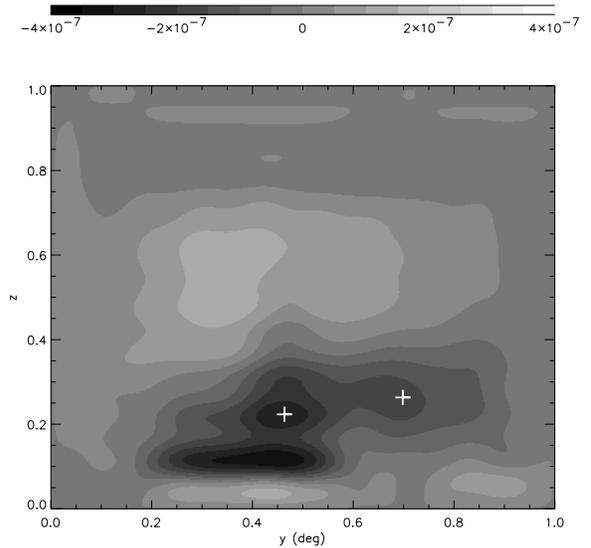,width=80mm}
\caption{Reconstructed gravitational potential for a notional
ground-based survey, after Wiener filtering; $n=30$ per sq arcmin,
$\gamma_{\rm rms} = 0.3$ per shear component. The clusters are
measured with $\nu=1.9$ and $\nu=1.4$.}
\end{figure}

On the other hand, we find that the full 3-D reconstruction of the
gravitational potential for typical cluster masses is difficult
even from space unless we include filtering (see Section 6.1). For example,
a measurement of the gravitational potential for a set of 25
clusters with mass $m=8\times10^{13}$ within 2 arcmin added together at
$z=0.25$, having smoothed $\phi$ in the radial direction with a
top hat with $\Delta z = 0.25$, produces a gravitational potential
amplitude for the central pixel of the cluster which is only 3.1
times larger than the rms noise. However, the noise amplitude for
larger $z$ grows rapidly (c.f. Section 6.1), making 3-D
measurement of a resolved cluster potential difficult without
filtering.

In order to improve on this, we can use the Wiener filtering described
in Section 3. We create a vector $\mathbf{P}$ containing our measured
gravitational potential along each line in the $z$ direction, and
measure the covariance of the $\Phi$ noise along this line of sight
from 100 zero-$\phi$ reconstructions of $\Phi$, recording this in a
matrix $\mathbf{N}$. We set the signal covariance matrix ${\mathbf
S}=(3\times 10^{-7})^2 \mathbf{1}$ where $\mathbf{1}$ is the unit
matrix, in order to select for a signal expected for a small cluster
mass (cf Figure 8). We then apply equation (\ref{eqn:wiener}) to our
gravitational potential vector for each line of sight.

Figure 10 shows the gravitational potential measurements for our
notional space-based survey after Wiener filtering (this is again a
reconstruction for the single field shown in Figure 8; after Wiener
filtering we no longer need to stack many fields to obtain a
signal). We measure the $z=0.25$ cluster with $\nu=4.2$ at the peak
(trough) pixel of its gravitational potential well, with $\nu=2.1$ at
the gravitational potential trough of the $z=0.4$ cluster. We also
find a substantial noise peak in the foreground at $z=0.1$. While the
recovery of the cluster gravitational potentials therefore constitutes
a challenging measurement, we are indeed able to reconstruct useful
information in the gravitational potential field itself. (The {\em
detection} significance of these clusters is much higher than the
measurement $\nu$ at a given point in the cluster; see Section 7 for
an approach to the detection significance.) Note the reduced absolute
amplitude of the gravitational potential in Figure 8; this is due to
the scaling in equation (\ref{eqn:wiener}).

\subsubsection{Gravitational Potential for Ground-based Experiment}

From the ground, we find that we can again reconstruct the lensing
potential, with $\nu \simeq 2.3$ for $0.75 < z < 1.0$ given a cell
size of 0.05 in $z$. As before, we could improve the signal-to-noise
by reducing our spatial resolution. However, reconstruction of the
gravitational potential itself (with Wiener filtering) is more
challenging than from space: Figure 11 demonstrates that we recover
the $z=0.25$ cluster with $\nu=1.9$ at its centre pixel, but have a
less significant measurement of the second cluster amplitude with
$\nu=1.4$ at its centre. The second cluster has the expected position
in $(x,y)$, but is offset to $z=0.3$; this degree of offset, $\Delta z
= 0.1$, is found to be typical for ground-based attempts at measuring
the 3D gravitational potential, due to the high $\Phi$ noise level
making Wiener filtering somewhat inaccurate in the $z$ direction. As
with our space-based experiment, we also find a substantial noise peak
in the foreground at $z=0.1$.

\section{Prospects for Mapping}

The results above are encouraging for the mapping of the 3-D lensing
fields. This includes the gravitational potential; we can use Wiener
filtering to detect individual mass concentrations, or can stack the
noisy potential field from many clusters in order to obtain
information on the typical gravitational profile of mass
concentrations. Here we will examine the prospects for mapping with
the various 3-D fields we have discussed so far, paying close
attention to the noise contributions to each field.

\begin{figure}
\psfig{figure=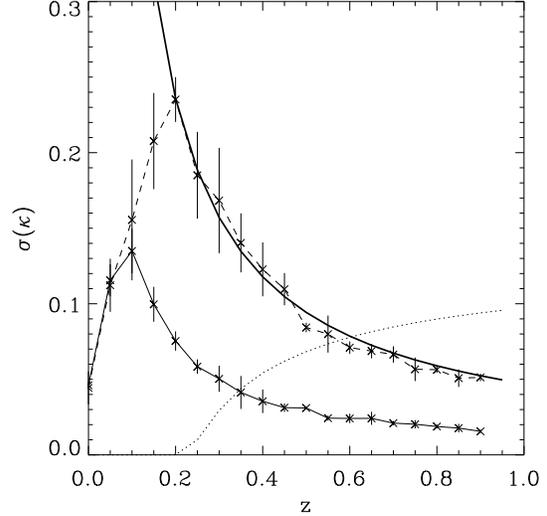,width=80mm}
\caption{Noise variance for 1 pixel (0.05 deg, 0.05 in $z$) as a
function of $z$ for the 3-D convergence field $\kappa$. The thin solid
line represents the measured noise level for our fiducial space-based
experiment, while the dashed line represents measured noise from our
ground-based experiment. The dotted line shows the expected signal
from a cluster at $z=0.2$. The thick solid line shows the noise
expected from theory for the ground-based survey, from
equation (\ref{eqn:kerr}); we see that there is close agreement between
simulations and theory.}
\end{figure}

\begin{figure}
\psfig{figure=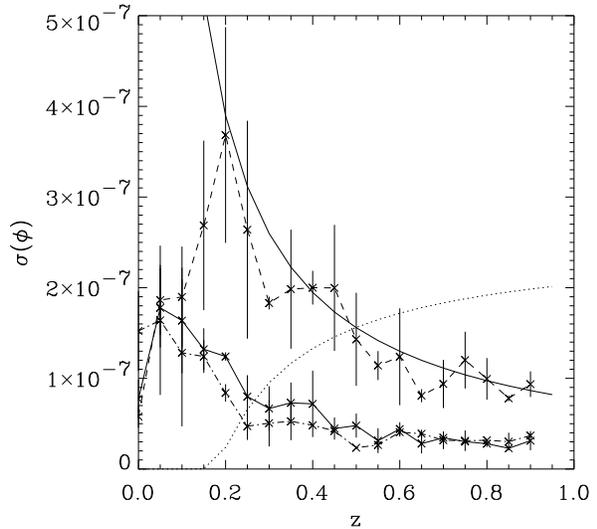,width=80mm}
\caption{Noise variance for 1 pixel (0.05 deg, 0.05 in $z$) as a
function of $z$ for the lensing potential in a 1 deg$^2$ square
survey. The solid line represents the measured noise level for our
fiducial space-based experiment, while the dashed line represents
measured noise from our ground-based experiment. The dash-dotted line
shows the effect of redistributing the galaxies according to equation
(\ref{nz}) for our space-based experiment. The dotted line shows the
expected signal from a cluster at $z=0.2$. The thick solid line shows
the noise expected from theory for the ground-based survey, from
equation (\ref{phi_err}); we again find good agreement between
simulations and theory.}
\end{figure}

\begin{figure}
\psfig{figure=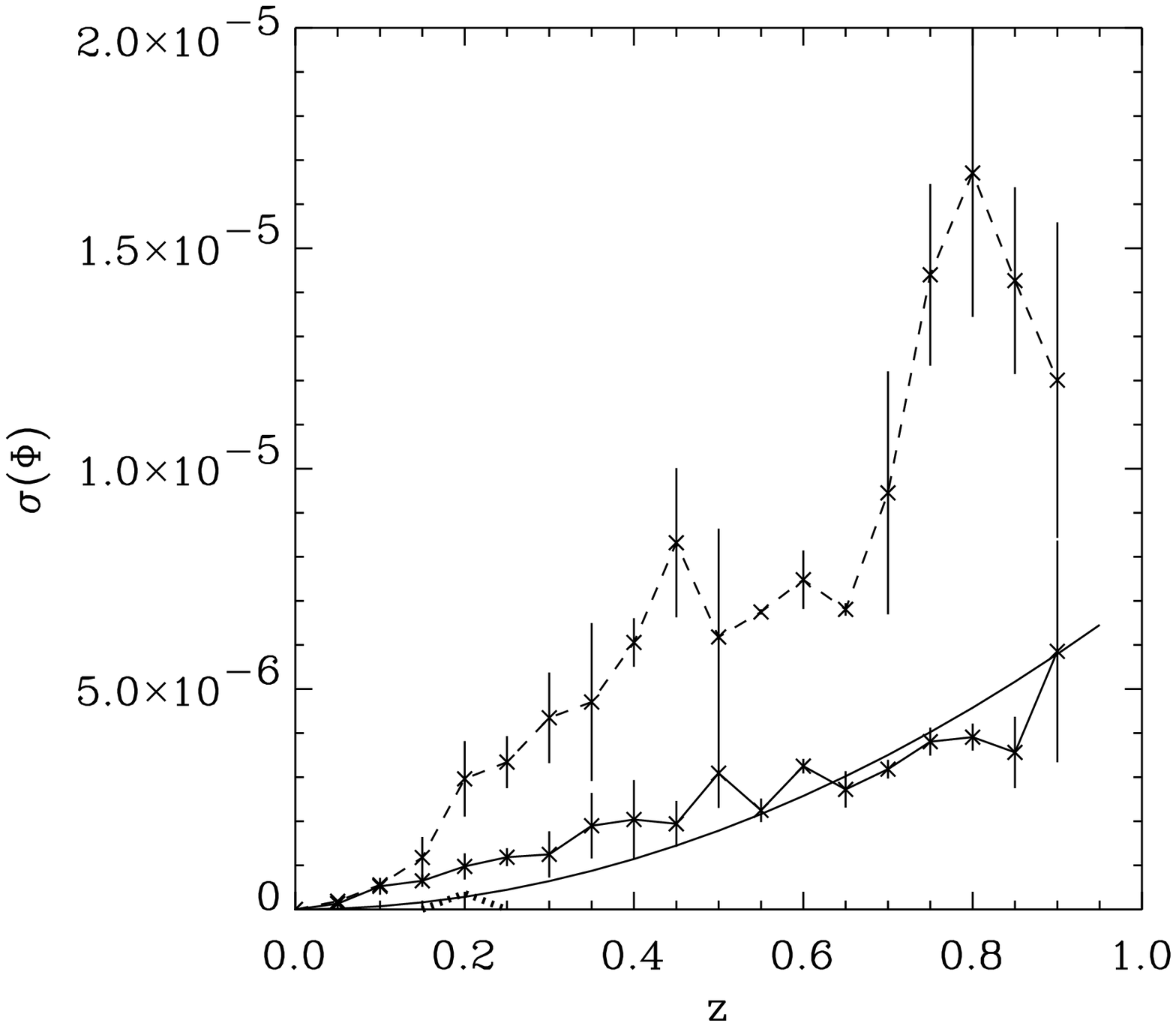,width=80mm}
\psfig{figure=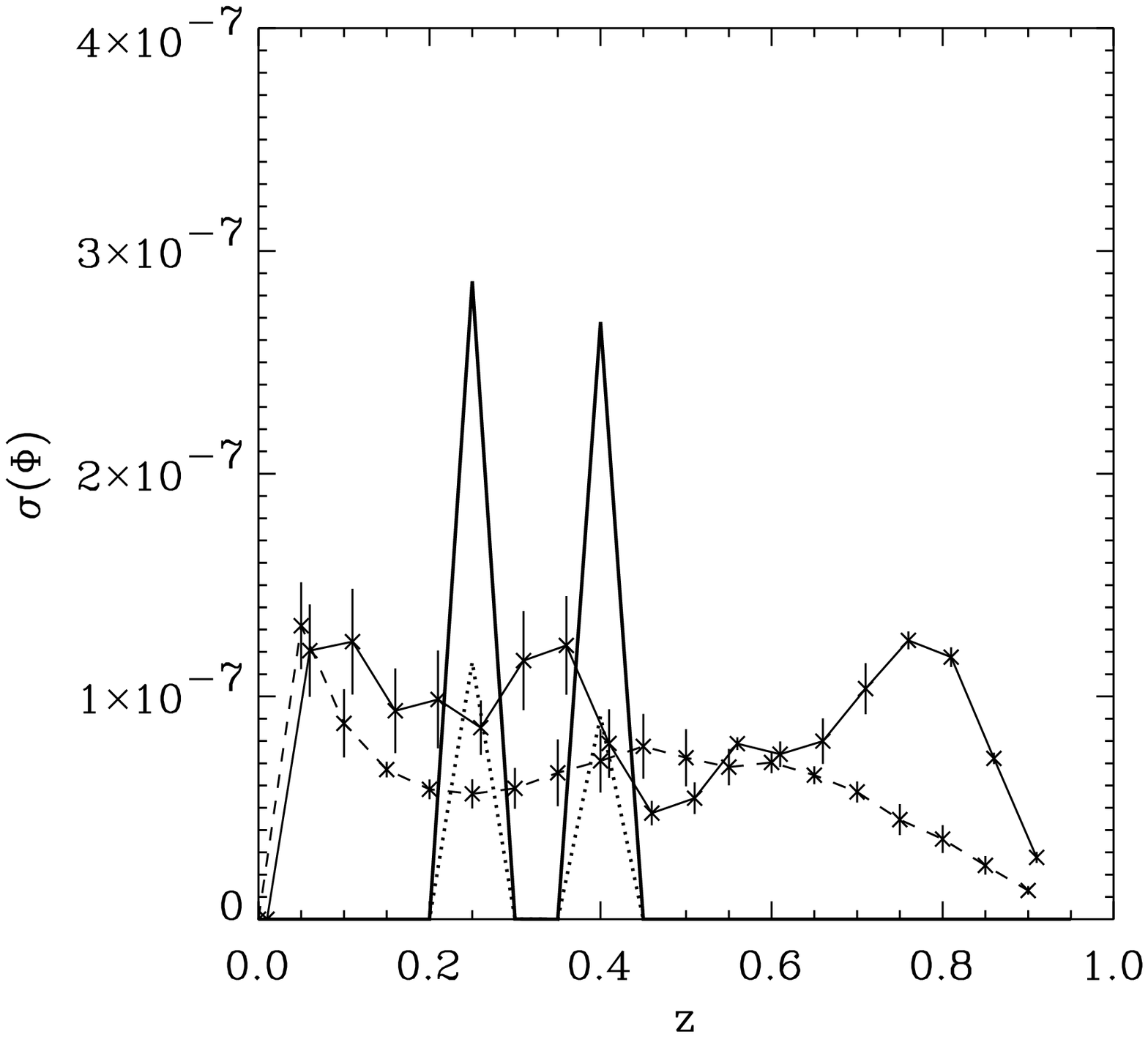,width=80mm}
\caption{Noise variance for 1 pixel (0.05 deg, 0.05 in $z$) as a
function of $z$ for the gravitational potential. The solid line
represents measured noise level for our fiducial space-based
experiment, while the dashed line represents measured noise from our
ground-based experiment. Top: standard gravitational potential
reconstruction. The dotted line shows the expected signal from a
cluster at $z=0.2$. The thick solid line shows the noise expected from
theory for the space-based survey, from equation (\ref{eqn:phierr});
we again find good agreement between theory and simulation. Bottom:
Wiener filtered reconstruction. The spikes show the measured
Wiener-filtered amplitude of a $10^{14}M_\odot$ cluster at $z=0.2$ and
$z=0.4$ for ground-based (dotted) and space-based (solid) noise
levels. }
\end{figure}

\subsection{Noise amplitudes}

In order to demonstrate the feasibility of our reconstructions, we
examine the noise from convergence, lensing potential and
gravitational potential maps measured in our simulations as a function
of $z$. Figure 12 shows the rms noise amplitude for $\kappa$ as a
function of redshift; Figure 13 shows the corresponding uncertainty in
$\phi$, while Figure 14 shows the uncertainty in $\Phi$. In each case,
we show the measured noise amplitude for space-based and ground-based
experiments, measured as a function of $z$, along with the expected
potential amplitude from a $m=10^{14}M_\odot$ cluster. Here we have
used the averaging scales described above, i.e. we have calculated the
noisy $\kappa$ and $\phi$ fields on a grid with total width 1 degree,
with grid spacing of 0.05 degree and 0.05 $c/H_0$ in $z$. It is simple
to derive noise amplitudes for other survey configurations from
equations (\ref{transwind}) and (\ref{eqn:phierr}).

We see from from the theoretical curves on Figure 12 that the noise
amplitude measured for the convergence field is in good agreement with
that calculated in Section 2. Similarly, we see that the measured noise
amplitude for $\phi$ in Figure 13 agrees well with the predicted
theoretical curve. Lastly, we see in Figure 14 that the uncertainty in
gravitational potential is in agreement with our theoretical model.

Note that, for the convergence and lensing potential, the noise
amplitude is such that $\nu\simeq 6$ measurement of these fields from
a typical $z=0.2$ cluster is possible in (0.05 degree, 0.05 in
redshift) bins in a space-based experiment, if we examine the
background potential at $z>0.5$ in the centre of the cluster. For a
ground-based experiment, the signal-to-noise in these bins is $\simeq
3$. Thus as we saw above, it is possible to map the properties of
clusters in terms of the lensing potential from space or ground. Note
the effect in Figure 13 of redistributing the galaxy distribution
according to equation (\ref{nz}); the amplitudes of the noise are
comparable, with a slight decrease in noise at low redshift and a
corresponding increase at high redshift for the galaxy distribution of
equation (\ref{nz}).

On the other hand, the gravitational potential without Wiener
filtering is only measured at $\nu=0.38$ in equivalent bins
around $z=0.2$ for space-based experiments, and has a signal-to-noise
of only $\simeq 0.12$ for ground-based measurements. Since the noise
reduces as $\sqrt{N}$ where $N$ is the number of stacked fields, we
would have to stack $\sim 40$ fields in order to resolve the
gravitational potential at the $\nu=2.5$ level, for a space-based
experiment without Wiener filtering. Increasing the size of bins is
not an option in this case, as we will lose spatial resolution for
examining the profile of the cluster.

However, Wiener filtering allows us to recover cluster masses
successfully, as shown in the bottom panel of Figure 14. Here we see
that, for a cluster mass of $10^{14} M_\odot$ at $z=0.25$, we obtain
$\nu=2.1$ measurements from the ground and $\nu=3.3$ measurements from
space. Note the differing noise and signal levels expected after
Wiener filtering for ground and space-based experiments; this is due
to the differing weightings in equation (\ref{eqn:wiener}) given
different input noise levels.

\subsection{Dependence on transverse and radial pixel sizes}

\begin{figure}
\psfig{figure=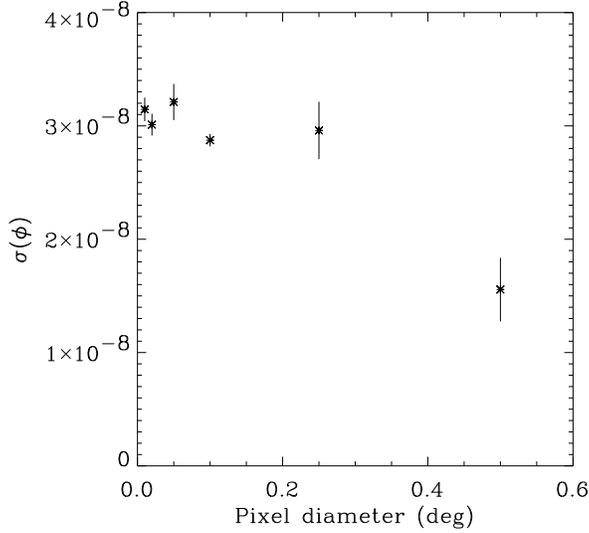,width=80mm}
\caption{Noise variance for the lensing potential at z=1 as a function
of angular pixel size, for our space-based experiment, with a 1 deg$^2$
square survey.}
\end{figure}

\begin{figure}
\psfig{figure=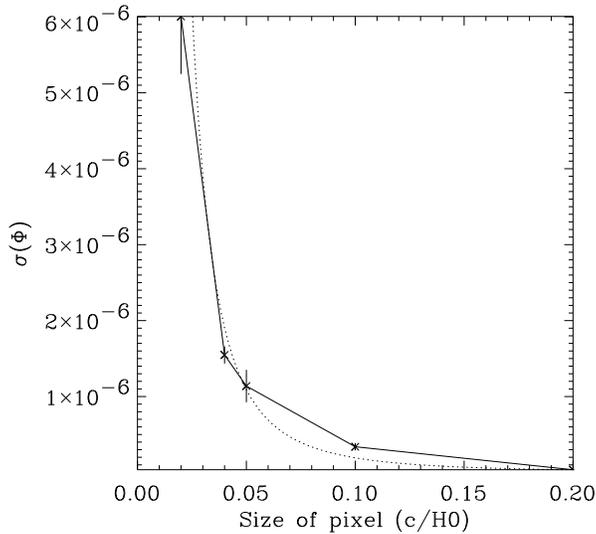,width=80mm}
\caption{Noise variance for the gravitational potential as a function
of pixel size in the radial direction, for our space-based experiment
(at z=0.4, 1 deg$^2$ survey, angular pixel diameter 3'). The dotted
line is the expected dependence from equation (\ref{eqn:phierr}).}
\end{figure}

The effect of increasing the angular pixel scale in a survey is of
interest, as this might be thought to increase
signal-to-noise. However, Figure 15 shows how an increase in pixel
size for a fixed survey size reduces the lensing potential noise
variance. We note that the effect is small until a pixel size $\sim
0.25 \Theta_{survey}$ is used; at this point, the pixel size will
usually be far too large to be of use to us, as we will typically be
interested in spatially resolving objects with the lensing
potential. The cause of the decrease in noise as pixel size increases
is partially the ln $r$ force law involved in the analysis in Section
3.1.2, and partially the increased bin size leading to averaging out
of the small-scale noise.

Figure 16 shows the effect of increasing the radial bin size for the
gravitational potential reconstruction. We see a reduction of the
noise level in agreement with equation (\ref{eqn:phierr}),
i.e. $\sigma(\Phi) \propto (\Delta z)^{-5/2}$. Thus if we are
unconcerned with radial resolution, we can increase our
signal-to-noise for gravitational potential by increasing $\Delta z$;
however, we will often be attempting to locate a mass concentration in
the radial direction, so this procedure should be used with caution.

\begin{figure}
\psfig{figure=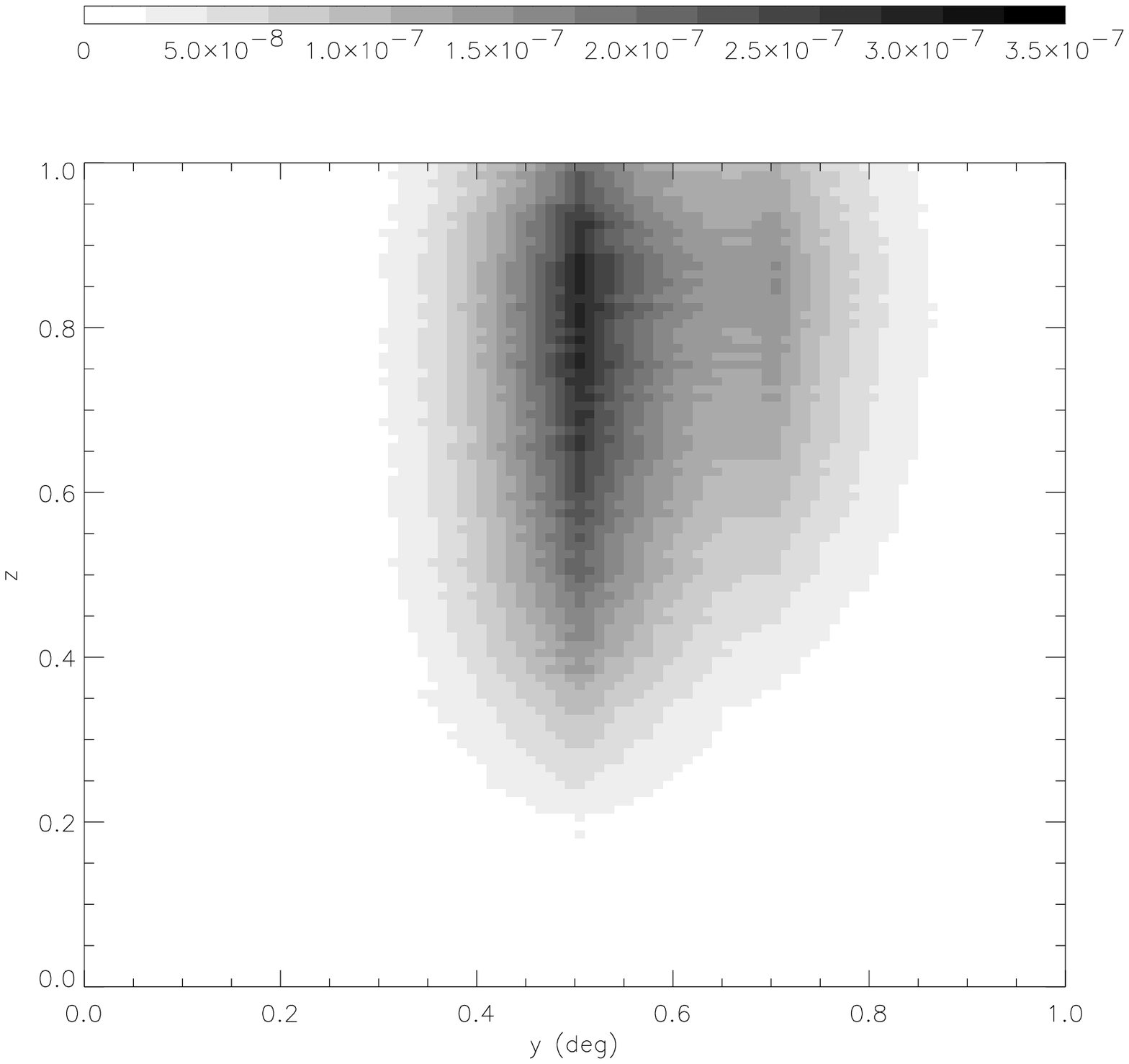,width=80mm}
\psfig{figure=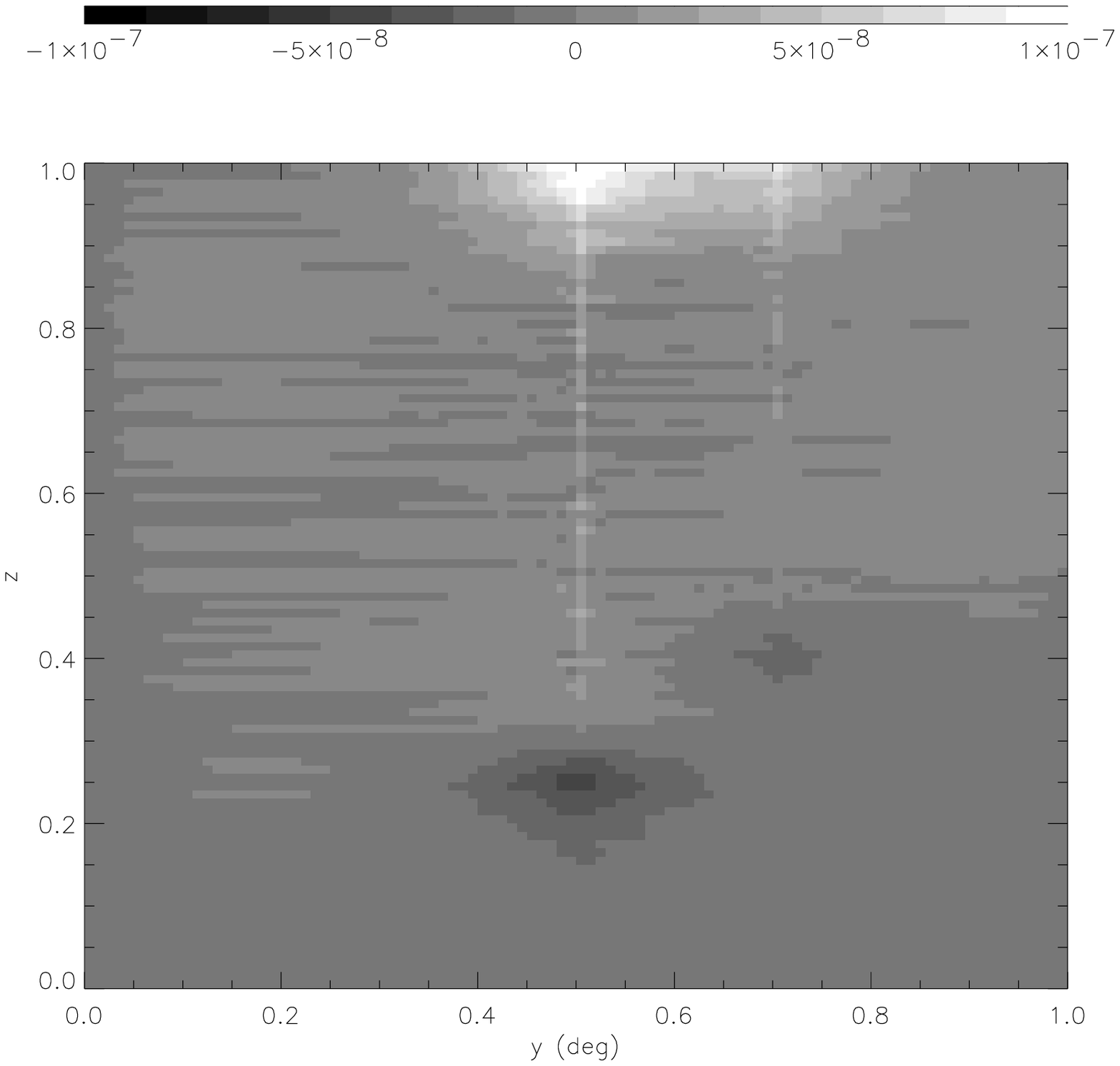,width=80mm}
\caption{Top panel: reconstructed lensing potential using the full
shear field of Figure 6, while including redshift uncertainty $\Delta
z = 0.1$. This is the usual $(y,z)$ plane at $x=0.5$. Bottom panel:
difference between input and recovered lensing potential fields.}
\end{figure}

\begin{figure}
\psfig{figure=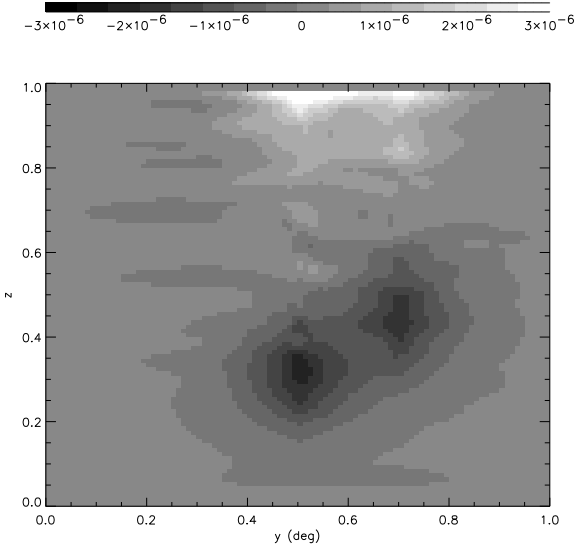,width=80mm}
\caption{Reconstructed gravitational potential using the full shear
field of Figure 6, including redshift uncertainty $\Delta z =0.1$.}
\end{figure}

\subsection{Redshift Errors}

In the above simulations, we have demonstrated that the Poisson noise
of sampling from a finite set of galaxies, together with the noise due
to galaxy ellipticities, represent serious sources of uncertainty for
our reconstruction, which must be overcome by averaging the signal
within sufficiently large pixels or filtering the signal. However,
there remains the further source of error due to redshift
measurement uncertainty, which we examine here with
our simulations (see Sections 3.4 and 3.5 for analytical discussion).

We can examine redshift errors with our simulations by allowing an
uncertainty in the redshift of each galaxy, in the following
fashion. For each galaxy, a $z$ value is drawn uniformly within the
$z$ slice in question; an uncertainty is introduced in this $z$
coordinate, by drawing from a Gaussian random variable with $1\sigma$
width $\Delta z = 0.1$ (pessimistically, for photometric redshifts; cf
Brown et al 2002 with $\Delta z = 0.05$ in $0<z<0.8$). If the new $z$
coordinate (resulting from adding this random offset to the original
$z$ position) is moved to a new shell, the galaxy (with its shear
calculated for the slice which it intially belonged to) is moved to the
neighbouring slice.

Figures 17 and 18 show the result of this process when the shear is
fully known everywhere; note that even with this large redshift
uncertainty, the change in the lensing potential reconstruction is
small ($\simeq 10$\% at redshifts near the cluster redshift, and less
elsewhere). Note that this smearing effect leads to slightly higher
lensing potentials in front of the cluster, and slightly slower rise
in the potential behind the cluster. Thus, for lensing potential
reconstructions, this effect will not dominate the noise. However,
Figure 18 shows that the result of such an uncertainty will be a
smearing of the cluster gravitational potential with smearing width $\simeq
\Delta z$; understandably, we cannot reconstruct the potential with a
resolution greater than our redshift resolution.

\section{3-D Information from the Lensing Potential}

\begin{figure}
\psfig{figure=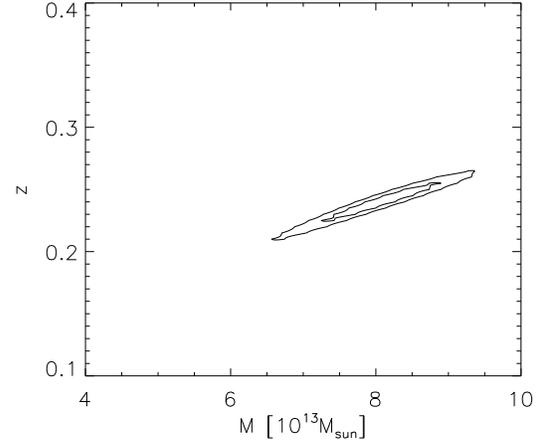,width=80mm}
\caption{1 and 2$\sigma$ $\chi^2$ fit constraints on $z$ and $m$ for 1
cluster along a line of sight, $m=8\times10^{13} M_\odot$,  with noise
appropriate for our space-based experiment. Here we are fitting the
simulated 3-D $\phi$ field with the $\phi$ field from a 2-D NFW
profile with parameters ($z,m$).}
\end{figure}

\begin{figure}
\psfig{figure=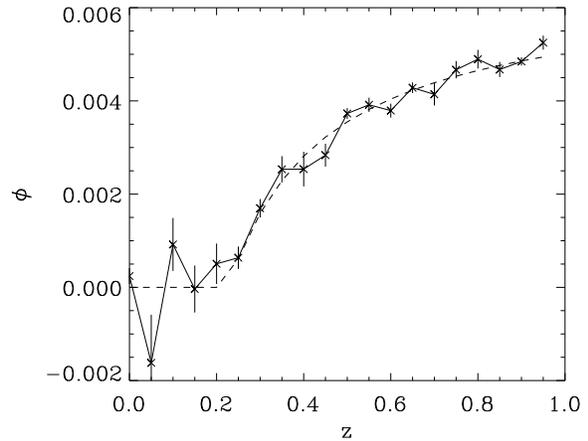,width=80mm}
\caption{Best fit model for $\phi$ field along line of sight through
centre of a cluster with $m=8\times10^{14} M_\odot$, with noise
appropriate for our space-based experiment.}
\end{figure}

It will be noted from our simulations that recovery of the
gravitational potential is more difficult than an adequate
reconstruction of the lensing potential. It is easy to see why this
should be so; the gravitational potential requires double
differentiation of the already noisy lensing potential
field. Therefore we are interested in both the lensing and
gravitational potentials; while the lensing potential is more easily
accessible, and itself contains useful topological information about
the mass field, the gravitational potential is the more fundamental
quantity.

A particular example of what is achievable by studying the $\phi$
field is the characterisation of the 3-D matter distribution on
cluster scales. Treating a cluster as a mass delta function in the $z$
direction (c.f. Hu \& Keeton 2002), it is clear from equation
(\ref{pot}) that the expected lensing potential in a flat universe due
to a cluster at radial position $r_s$ will be

\begin{equation}
\phi(z) = \left\{
  \begin{array}{ll}
    0               & r \le r_s \\
    2\rho dV \frac{r-r_s}{r r_s} & r > r_s \\
  \end{array}\right. 
\label{phidelta}
\end{equation}
where $\rho dV$ is the mass content of the source pixel.

We can use a superposition of such cluster contributions to fit a
given $\phi$ field, and thus discover significant clusters behind
clusters, for example (c.f. Hu \& Keeton 2002). We can also directly
measure constraints on mass and position of clusters without using the
redshift of the cluster members themselves (e.g. Wittman et al 2001, 2002).

\subsection{Single cluster $\chi^2$ fitting}

In order to demonstrate these applications, we first simulate a
cluster at $z=0.25$ with mass $8\times 10^{13} M_\odot$ within a 2
arcmin radius using the simulation recipe described in Section 3,
including realistic galaxy distribution and ellipticity (using our
space-based parameters, $n=100$, $\sigma_\gamma=0.2$ per
component). After measuring the resulting $\phi$ field as in Section
4, we applied a $\chi^2$ fitting procedure for the mass $m$ and
position $z$ of the cluster. This was achieved by setting up a 2-D NFW
profile at radial position $z$, normalised to mass $m$; the expected
3-D $\phi$-field for this profile was calculated according to equation
(\ref{phidelta}). $\chi^2$ for the data with respect to this profile
was calculated for an array of $m$ and $z$ values, with step size
$2.5\times 10^{11} M_\odot$ in $m$ and 0.005 in $z$. Note that the
$x,y$ position and radius of our NFW test profile were fixed to the
actual position and radius of the cluster in this experiment; in a
practical scenario one would apply a $\chi^2$ fit for these parameters
as well, or infer them from the galaxy positions of cluster members.

Figure 19 shows the resulting constraints on mass and radial position;
we find a highly significant detection of the cluster at the
$\Delta\chi^2=186$ level, thus we can certainly use this 3-D approach
to detect at least one cluster along the line of sight. We obtain
accurate measurements of the mass ($m=(8.07\pm0.83)\times 10^{13}
M_\odot$ within 2' radius) and position $(z=0.24\pm 0.016)$,
which makes this approach promising for examining clusters in
3-D. Figure 20 demonstrates a best-fit $\phi$ field for this form of
simulation, with mass multiplied by 10 for illustrative purposes.

\begin{figure}
\psfig{figure=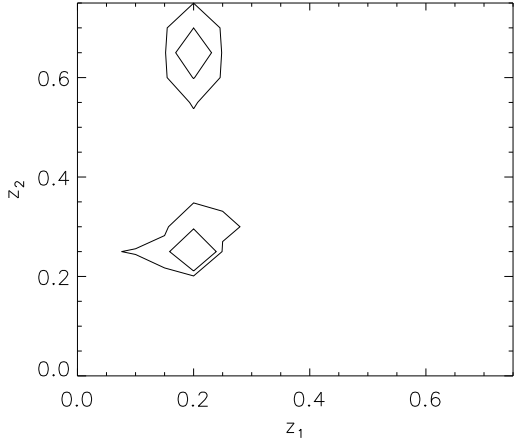,width=80mm}
\caption{1 and 2$\sigma$ $\chi^2$ fit constraints on $z_1$ and
$z_2$ for 2 clusters along line of sight, with noise appropriate for
our space-based experiment, having marginalised over the masses of
the two clusters.}
\end{figure}

\begin{figure}
\psfig{figure=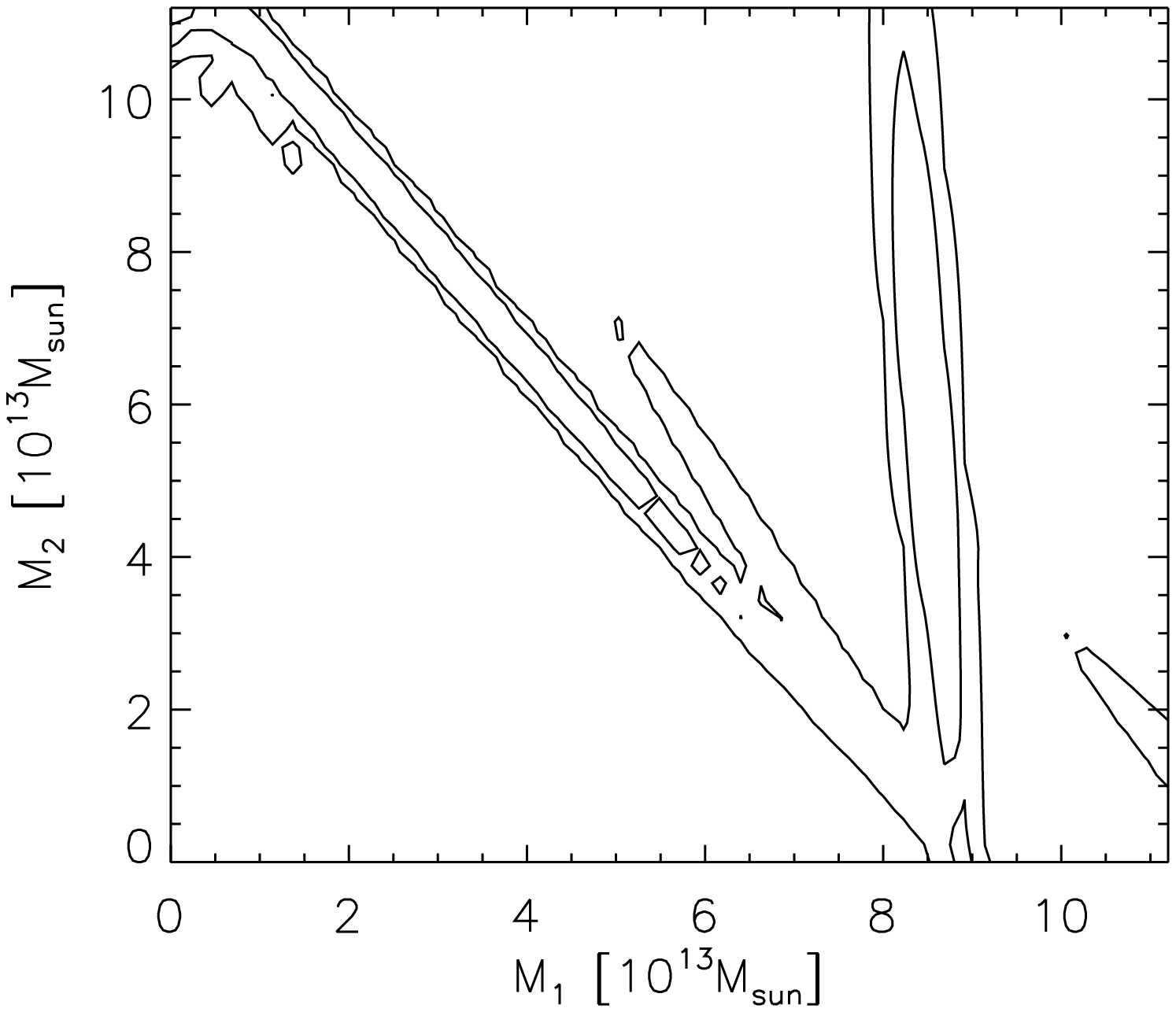,width=80mm}
\caption{1 and 2$\sigma$ $\chi^2$ fit constraints on $m_1$ and
$m_2$ for 2 clusters along line of sight, with noise appropriate for
our space-based experiment, having marginalised over the positions of
the two clusters.}
\end{figure}

\begin{figure}
\psfig{figure=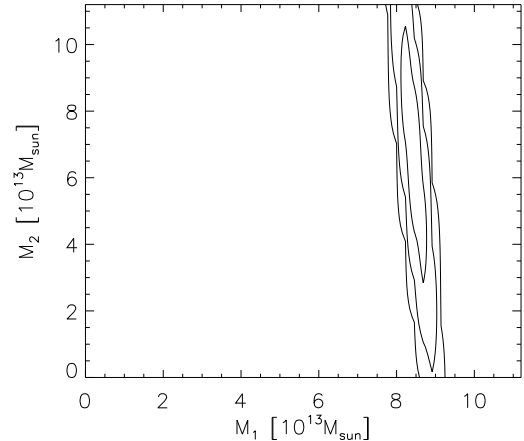,width=80mm}
\caption{1, 2 and 3$\sigma$ $\chi^2$ fit constraints on $m_1$ and
$m_2$ for 2 clusters along line of sight, with noise appropriate for
our space-based experiment, after including redshift measurements of
cluster positions.}
\end{figure}

\begin{figure}
\psfig{figure=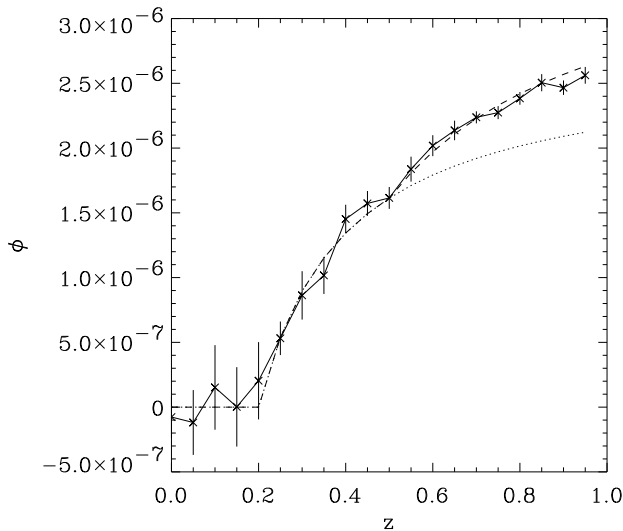,width=80mm}
\caption{Best fit model for $\phi$ field along line of sight through
centre of 2 clusters, $m=8\times10^{14} M_\odot$, at $z=0.25$ and
0.5. The dotted line represents the $\phi$ field arising from the
$z=0.25$ cluster alone.}
\end{figure}

\subsection{$\chi^2$ fitting for two clusters along the line of sight}

We can examine the possibility of detecting clusters behind clusters
by simulating an $8\times10^{13}M_\odot$ cluster (2 arcmin radius) at
$z=0.6$ behind another $8\times10^{13}M_\odot$ cluster at $z=0.25$,
and applying a $\chi^2$ fit for two masses and positions using the
method above.

Figures 21 and 22 show the constraints we obtain on mass and position
of the two clusters from the simulation, having marginalised
over the 4-dimensional $\chi^2$ distribution to find a subset of
parameters. We see that, for this example, two configurations of
clusters are possible: (a) two clusters very near each other in
redshift; this is essentially the discovery of a 1-cluster
solution. (b) One cluster at $z=0.25$ with another of similar mass at
$z=0.6$; this is the solution corresponding to our input scenario. In
this case, we obtain greater accuracy in measuring the mass for the
nearer cluster. This can be understood from Figure 24 (displaying
masses 10 times larger for illustrative purposes): much of the
background $\phi$ amplitude is due to the first cluster, so the mass
and position of this cluster is well-constrained; on the other hand,
the second cluster's $\phi$ contribution has less distance in which to
rise, so this cluster's mass and position estimates are more affected
by the noise amplitude.

If this double solution were present when the fitting procedure was
applied to real data, we could easily use our sample redshifts to
confirm one scenario by looking for increased number-counts at the
claimed cluster redshifts. By confirming the redshift of the cluster
in this fashion, we can obtain better estimates of the mass, together
with a more conclusive statement as to whether there are mass
concentrations in the background. This is demonstrated in Figure
23. Here we have again simulated a cluster behind a cluster; we have
now allowed the clusters to have known redshifts in our $\chi^2$
fit. In this case we find that a two-cluster fit is substantially
better than a one cluster fit (i.e. $m_2 = 0$) at the $2\sigma$
level. We find best values for the masses of the clusters to be
$m_1=(8.46^{+0.31}_{-0.35})\times 10^{13}M_\odot$ and
$m_2=(6.63^{+3.92}_{-3.79})\times 10^{13}M_\odot$ (c.f. input
$m_{1,2}=8\times 10^{13}M_\odot$) within 2 arcmin radii. If we only
constrain $z_1$, we find a best fit value for the background position
of $z_2=0.6\pm0.05$ (c.f. input $z=0.6$); thus if we have found a
low-redshift cluster, we can check if there are significant clusters
behind it.

We could increase the number of clusters in our $\chi^2$ fit, in order
to seek for $>2$ clusters along the line of sight. However, in this
case we will begin to fit the noise rather than real structures; this
will be seen as too good a fit in our $\chi^2$ (i.e. $\Delta \chi^2
\la \sqrt{2 (x/\Delta x)(y/\Delta y)(z/\Delta z)}$). This restricts
our ability to construct accurate 3-D maps using this method; however,
we can at least map up to the first few significant mass concentration
in the $z$ direction, and cosmological information could be gained by
determining the statistical properties of the distance to the first or
first few mass concentrations.

One can envisage, therefore, a procedure consisting of (a) initial
detection of mass concentrations using the 3-D distortion field alone,
followed by (b) improved mass and background structure estimates by
assigning accurate redshifts from the visible matter associated with
the detected foreground mass concentrations.

\section{Conclusions}

In this paper, we have developed and tested a practical method for 3-D
reconstruction of the gravitational potential via weak lensing
measurements, together with the more easily obtained lensing
potential. This methodology is based on the reconstruction equations
of Kaiser \& Squires (1993) and Taylor (2002), by which these local
3-D potentials can be calculated given knowledge of the lensing shear
field in 3-D. This can be obtained by using shear estimators for
galaxies with known redshifts.

We have presented analytical forms for the shot-noise uncertainty
in the convergence, lensing potential, gravitational potential and
density field, noting that these fields become progressively more
noisy for realistic survey sizes. We have also calculated the
effects of only having finite sky-coverage for a survey and
estimated the variance measured on such a survey, and the
additional uncertainty in the reconstruction due to structure
beyond the survey boundary. In particular we have found that the
contribution to the reconstruction uncertainty in the differential
lensing potential, $\Delta \phi$, is dominated by large-scale
structures.

We have also shown that further sources of error, including
photometric redshift errors, redshift-space distortions, and
multiple scatterings of light rays, will not be dominant in our
reconstruction process.

In order to simulate the measurement of a 3-D gravitational field, we
have calculated the expected lensing potential due to a given mass
field upon a 3-D grid. From this we have calculated the shear expected
upon a galaxy image for a galaxy positioned anywhere in the 3-D
grid. We have produced a catalogue of galaxies positioned in
accordance with a realistic redshift distribution and given each an
appropriate shear plus a random intrinsic ellipticity; this final
galaxy shear catalogue was the information given to our reconstruction
software.

We have made reconstructions of the lensing and gravitational fields,
by smoothing the noisy shear data, calculating the lensing potential
according to Kaiser \& Squires (1993) and using Taylor's equation
(\ref{inv}) to find the gravitational potential.

We have found that the method works well in reconstructing the full
lensing and gravitational potentials in the absence of noise. However,
the addition of Poisson sampling of the field at a finite set of
galaxy positions, together with the noise due to the intrinsic
ellipticity of objects, produces significant sources of error. We find
that we obtain lensing potential maps with $\nu \simeq 6$ in [3', 3',
0.05] bins in angle and redshift for a cluster of mass $8\times
10^{13}M_\odot$. Unfortunately, corresponding gravitational potential
maps have only $\nu \simeq 0.5$ in pixels of this size. However,
applying Wiener filtering to this gravitational potential, we can
obtain $\nu \simeq 3$ measurements of gravitational potential of
clusters of mass $8\times 10^{13}M_\odot$. 

This provides excellent prospects for obtaining cosmologically
significant information directly from the measured gravitational
potential field. For surveys with a redshift limit $z=1$, mass
concentrations $\ga 10^{14}M_\odot$ can be directly mapped between $0.1 \la z
\la 0.5$, while statistical mapping can be used to examine the
gravitational potential fluctuations on smaller mass scales.

We have examined the effect of redshift uncertainties upon our
simulations, and find that these contribute a much smaller error than
the dominant intrinsic ellipticity noise term.

Finally we have emphasised that, even on scales where a gravitational
potential measurement is uncertain, the 3-D measurement of the $\phi$
field is valuable, and can give us useful information about the
statistics and topology of the mass field. In particular, we can
obtain accurate measurements of the mass and position of clusters
along the line of sight, and significantly detect the presence of
clusters behind clusters.

The methodology described here for engaging in 3-D gravitational
mapping has numerous applications. We can obtain direct measurements
of mass distributions in 3-D, which can act as an important
cosmological probe via the mass function or cluster number counts. We
can directly measure cross correlation functions between mass and
light in 3 dimensions. Also, for possible 'dark' mass concentrations
(e.g. Erben et al 2000), this 3-D mapping procedure allows us to
measure both mass and radial position for such objects, which is quite
impossible via conventional redshift methods.

\section{Appendix: Calculating uncertainties on the 3D fields}

Here we describe the details of our analysis for calculating
uncertainties on the lensing and gravitational potentials.

 \subsection{The lensing potential field}

In Section 3.1.2 we state the covariance of the 3-D lensing potential,

\be
\lgl \phi(\r) \phi(\r') \rgl_{SN} = 4 \partial^{-2} \partial'^{-2}
 \lgl \kappa(\r) \kappa(\r') \rgl_{SN}.
 \label{phicov1}
\ee
We can calculate this covariance over the observed area, assuming a
flat sky, obtaining
\ba
    \lefteqn{\lgl \phi(\r) \phi(\r') \rgl_{SN} =}\nn
 & &    \frac{\gamma_{\rm rms}^2}{\pi^2 n(r)r^2}
        \int_A d^2 \theta'' \ln |\thetab - \thetab''| \ln |\thetab'  -
        \thetab''| \delta_D(r-r')
        \label{phicovappend}
\ea
where the integral is taken over the survey area, $A$. The discrete
case is an obvious change to a summation over bins.

Transforming to the differential potential field,
\be
    \Delta \phi = \phi - \overline{\phi},
\ee
where $\overline{\phi}$ is the mean field estimated from a finite
survey of area $A$,
\be
    \overline{\phi} = \frac{1}{A} \int_A \! d^2 \! \theta \, \phi
    (\thetab),
\ee we find the covariance $\lgl\Delta \phi(\r) \Delta
\phi(\r')\rgl_{SN}$ is equivalent to equation (\ref{phicovappend})
after transforming the kernel \be
    \ln |\thetab - \thetab'| \rightarrow \ln | \thetab - \thetab'| -
    \frac{1}{A} \int_A \! d^2 \! \theta \,\ln |\thetab - \thetab'|.
\ee
In the simple case of a circular survey with radius $R$ we find
\be
    \frac{1}{A} \int_A \! d^2 \! \theta' \, \ln |\thetab - \thetab'| =
     \frac{1}{2} [(\theta/R)^2-1 + \ln R^2].
\ee
Note here we have assumed infinite resolution for the survey, or
infinitely small pixels.

The uncertainty in the 3-D lensing potential difference is then given by
\be
    \lgl \Delta \phi^2 (\r) \rgl_{SN} =
    \frac{\gamma_{\rm rms}^2}{\pi^2 n(r)}
    \frac{\Theta^2(\thetab)}{r^2}  \delta_D(r-r'),
\ee
where
\be
    \Theta^2(\thetab)=\int_A \! d^2\! \theta'\, \left[\ln |\thetab - \thetab'|-
    \frac{1}{A} \int_A \! d^2 \! \theta \, \ln |\thetab - \thetab'|\right]^2,
\ee
which in general has to be evaluated numerically. In the special case
of $\thetab=0$ and a circular aperture with angular radius $\theta$,
this can be evaluated analytically, giving
\be
    \lgl \Delta \phi^2(\thetab = 0) \rgl_{SN} = \frac{5}{24 \pi}
  \frac{\gamma^2_{\rm rms}}{n(r)} \frac{\theta^2}{r^2} \delta_D(r-r'),
\ee
where we have taken into account the conical geometry of the
survey. This is the result discussed in Section 3.1.2.

\subsection{The Newtonian potential field}

In order to calculate the uncertainty on the 3-D Newtonian potential,
we must smooth the field; this is because we require a double
differentiation of the lensing potential, but only sample this field at
discrete points where there are galaxies. If we smooth in the radial
direction with an arbitrary smoothing kernel, $w(r)$, we find the
resulting covariance matrix of the Newtonian potential for a constant
galaxy number density, $n$, in the distant observer approximation, is
\be
\lgl \Phi(\r) \Phi(\r') \rgl_{SN} = \frac{\gamma^2_{\rm rms}}{4 \pi^2 n }
\frac{\Theta^2(\thetab,\thetab')}{L^3(r,r')}, \ee where \be
\frac{1}{L^3(r,r')} = \frac{r^2 r'^2}{R^2}\int \! dy \, w''(r,y)\,
w''(r',y)
\ee
has units of inverse volume. Dashes on the window function denote
derivatives with respect to distance. If the number density of sources
is not a constant, these formulae must be altered by the substitution
\be
    \frac{1}{nL^3(r,r')} \rightarrow \frac{r^2 r'^2}{R^2}  \int \! dy \, w''(r,y)
        w''(r',y)/ n(y).
\ee
Again we can evaluate these expressions for the variance in a bin when
$r'=r$ and assuming a Gaussian weighting function, $w(r) = [\sqrt{2
\pi} r_{||}]^{-1} \exp( -r^2/2r_{||}^2)$, with smoothing radius
$r_{||}$. In this case to leading order, when $r \gg r_{||}$, and
taking into account the conical geometry of the survey, the
uncertainty on the Newtonian potential conveniently reduces to
\be
    \lgl \Phi^2(r) \rgl_{SN} =  \frac{5}{64 \pi \sqrt{2 \pi}}
    \frac{\gamma^2_{\rm rms}}{r_{||}^3 n}\left( \frac{r}{r_{||}}\right)^4
    \left(\frac{r_{||}}{R} \right)^2 \theta^2 .
\ee
This is the result discussed in Section 3.1.3.

\section*{Acknowledgments}

DJB is supported by a PPARC Postdoctoral Fellowship; ANT is supported
by a PPARC Advanced Fellowship. We thank Martin White, Alan Heavens,
Meghan Gray and Simon Dye for very useful discussions.

\label{lastpage}

\end{document}